\documentclass[a4paper,fleqn]{cas-sc}

\usepackage[numbers,sort&compress]{natbib}
\usepackage{subcaption} 
\usepackage{xcolor} 
\usepackage{amssymb} 
\usepackage{pifont}  

\newcommand{\cmark}{\ding{51}} 
\newcommand{\xmark}{\ding{55}} 

\def\tsc#1{\csdef{#1}{\textsc{\lowercase{#1}}\xspace}}
\tsc{WGM}
\tsc{QE}

\newtheorem{theorem}{Theorem}
\newtheorem{lemma}{Lemma}
\newtheorem{assumption}{Assumption}
\newdefinition{rmk}{Remark}
\newproof{pf}{Proof}

\begin{document}
\let\WriteBookmarks\relax
\def\floatpagepagefraction{1}
\def\textpagefraction{.001}

\shorttitle{}    

\shortauthors{}  

\title [mode = title]{A Model-Based Extended State Observer for Discrete-Time Linear Multivariable Systems}

\author[1]{Jinfeng Chen}



\ead{particlefilter2012@gmail.com}



\affiliation[1]{organization={Technology Division},
            addressline={University of Houston}, 
            city={Houston},
            postcode={77004}, 
            state={TX},
            country={USA}}

\author[2]{Zhiqiang Gao}


\ead{z.gao@csuohio.edu}



\affiliation[2]{organization={Center for Advanced Control Technologies},
            addressline={Cleveland State University}, 
            city={Cleveland},
            postcode={44115}, 
            state={OH},
            country={USA}}

\author[1,3]{Qin Lin}

\cormark[1]


\ead{qlin21@central.uh.edu}



\affiliation[3]{organization={Department of Electrical and Computer Engineering},
            addressline={University of Houston}, 
            city={Houston},
            postcode={77004}, 
            state={TX},
            country={USA}}

\cortext[1]{Corresponding author.}


\begin{abstract}
In the absence of a detailed and accurate plant model, extended state observer (ESO) has been clearly shown as an effective tool in both control applications and theoretical analysis. Less clear is how to further enhance such a tool when the model is available, however inaccurate, together with additional measurement information. To this end, a model-based ESO (MB-ESO) and its generalized form (GMB-ESO) for discrete-time linear multivariable systems are developed in this paper by explicitly incorporating the additional state-space model and measurement information, which is otherwise bypassed by the conventional ESO. In addition, GMB-ESO is shown to accommodate systems with non-diagonal disturbance gain matrices. {Necessary and sufficient conditions for the existence of the MB-ESO and GMB-ESO are established in terms of a well-defined disturbance vector relative degree and the absence of invariant zeros in the disturbance-to-measurement path. Under noise-free measurements and infinite observer bandwidth, exact disturbance reconstruction can be achieved within a finite number of steps, while under measurement noise, \(\varepsilon\)-accurate reconstruction can be guaranteed with finite observer bandwidth.} Finally, the disturbance {recontruction} error of the MB-ESO is shown, in both theoretical analysis and simulation, to decrease monotonically with time, when the disturbance gain matrix is diagonal, and an explicit bound on the disturbance estimation error is derived.
\end{abstract}


\maketitle

\section{Introduction}\label{sec:introduction}
{The reconstruction\footnote{{Reconstruction denotes deterministic recovery of disturbance from measurements, not stochastic estimation of its distribution~\cite{ansari2019deadbeat}}} of unknown disturbances has received sustained attention in control theory and applications. Unknown input observers (UIOs) provide a deterministic framework for reconstructing arbitrary unknown disturbances by decoupling them from the observer error dynamics. Their existence, synthesis, and reconstructability have been extensively studied through algebraic and geometric formulations, leading to well-established necessary and sufficient conditions based on rank matching and stable invariant zeros \cite{basile1969on,hou1992design,hou1998input,hautus1983strong,willems1981disturbance}. Subsequent developments, including delayed UIOs and high-gain geometric formulations, have relaxed structural constraints or generalized exact decoupling by using past measurement histories or approximate decoupling \cite{sundaram2007delayed,willems1981almost,willems1982almost}. Overall, UIO theory provides a rigorous reconstructability framework for arbitrary disturbances.

In parallel, a broad class of practical augmented system observers (ASOs) has been developed to jointly reconstruct system states and disturbances in the presence of measurement noise. Originating from the idea of augmenting disturbances as additional states in the late 1960s \cite{johnson1968optimal}, this family includes extended high-gain observers (EHGOs) \cite{freidovich2008performance}, extended state observers (ESOs) \cite{han2009from}, generalized ESOs (GESOs) \cite{li2012generalized}, proportional-integral observers (PIOs) \cite{soffker1995state}, and generalized PIOs (GPIOs) \cite{sira2013on}. Despite substantial methodological progress and widespread applications, a fundamental research gap remains: the reconstructability theory of ASOs is still underdeveloped. In particular, compared with the well-established theory for UIOs, it is less systematically understood \emph{when} and \emph{why} arbitrary disturbances are reconstructable by ASOs and \emph{what} reconstruction capability can be achieved.

We further elaborate on this gap by highlighting three common limitations in this ASOs family. (i) chain-of-integrator forms. EHGOs, ESOs, and GPIOs require chain-of-integrator forms, in which the disturbance enters an input-equivalent channel that supports matched disturbance rejection. (ii) Restrictive disturbance assumptions. PIOs \cite{soffker1995state} and GESOs \cite{li2012generalized} extend beyond chain-of-integrator structures to systems with disturbances entering non-input channels but impose restrictive asymptotic assumptions: PIOs \cite{soffker1995state} require the disturbance to have a finite steady-state limit, whereas GESOs \cite{li2012generalized} explicitly assume convergence to a constant disturbance. These assumptions exclude many continuously varying or abruptly changing real-world disturbances. (iii) High observer gains make ASOs highly sensitive to measurement noise.

This paper aims to fill this research gap by establishing a theoretical foundation for disturbance reconstructability and addressing the three common limitations described above. To provide a clear taxonomy of these observer paradigms, Table~\ref{tab:relatedWork} compares several existing methodologies and highlights the positioning of our work. Our proposed model-based ESO (MB-ESO) and its generalized variant, the GMB-ESO, establish reconstructability conditions for arbitrary disturbances without imposing prior assumptions on disturbance behavior, while exploiting multiple measurements and model information to reduce the required observer gain and thereby improve robustness to measurement noise.
}
\begin{table*}[htbp]
{ 
    \caption{Comparison of the existing measurement-driven disturbance estimation methodologies.}
    \label{tab:relatedWork}
    \centering
    \begin{tabular}{cccccc}
        \toprule
        Methods & Existence Condition & \makecell{Bounded- \\disturbance- \\ and-derivative \\ Assumption} & \makecell{Disturbance \\Reconstruc- \\tability} & \makecell{Uncertainty in \\ Observer Error\\ Dynamics} \\
        \midrule
        PIO \cite{soffker1995state} & The augmented system is observable. & \cmark & \xmark & \cmark \\
        \hline
        GESO \cite{li2012generalized} & The augmented system is observable. & \cmark  & \xmark & \cmark \\ \hline
        ESO \cite{han1995extended} and GPIOs \cite{sira2013on} & \makecell{The system can be written \\in a cascade integral chain.} & \cmark & \cmark & \cmark\\
        \hline
        EHGO \cite{freidovich2008performance} & \makecell{The nonlinear system is in a\\ normal form with minimum phase.} & \cmark & \cmark & \cmark\\
        \hline
        UIO \cite{hautus1983strong,hou1994disturbance} & \makecell{1) A matrix rank matching condition;\\ 2) A stable invariant zeros condition.} & \xmark & \cmark & \xmark\\
        \hline
        D-UIO \cite{sundaram2007delayed} & A stable invariant zeros condition. & \xmark & \cmark & \xmark\\
        \hline
        \makecell{Ours\\(MB-ESO \& \\ GMB-ESO)} & \makecell{No invariant zeros condition \\(with dynamic extension \cite{isidori1995nonlinear, wu2020performance}).} & \xmark & \cmark & \cmark \\
        \bottomrule
    \end{tabular}
    }
\end{table*}



The main contributions of this paper are as follows:

{

\textbf{1) Reconstructability Conditions.} We establish necessary and sufficient conditions for the existence of the MB-ESO and GMB-ESO in reconstructing arbitrary disturbances without imposing assumptions on disturbance behavior. Under noise-free measurements, arbitrary disturbances can be reconstructed exactly within a finite number of steps with infinite observer bandwidth. In the presence of measurement noise, arbitrary disturbances can be reconstructed with \(\varepsilon\)-accuracy using a finite observer bandwidth. This contribution directly addresses the fundamental research gap identified above.

\textbf{2) Novel Observer Design.} Guided by the proposed reconstructability conditions, the MB-ESO and GMB-ESO provide a new pathway for exploiting multiple measurements and model information to reduce the required observer gain and thereby improve robustness to measurement noise. The MB-ESO is designed for systems with a diagonal disturbance gain matrix, where each disturbance can be reconstructed independently. The GMB-ESO extends this design to non-diagonal cases, where disturbances are mixed across measurement channels. It first aligns the channel-dependent delays and then separates the mixed reconstructions to recover the individual physical disturbances.

\textbf{3) Explicit Disturbance Reconstruction Error Bound.} To the best of our knowledge, this is the first work to establish an explicit bound on the disturbance reconstruction error by deriving its exact transfer function \cite{chen2023robust}. The resulting bound is nonconservative and shows that the reconstruction error depends only on the disturbance relative degree, observer eigenvalues, and disturbance variation, while remaining independent of other system properties.

\textbf{4) Monotonic Disturbance Reconstruction Error.} We establish a complementary transient property of the disturbance reconstruction error. When the disturbance gain matrix is diagonal, the reconstruction error of the MB-ESO decreases monotonically over time, ensuring that no overshoot or oscillation occurs during disturbance reconstruction.

\textbf{5) Effect of Discretization on Disturbance Reconstruction.} Although discretization is required for the hardware implementation of ESOs \cite{miklosovic2006discrete}, it can introduce invariant zeros between disturbances and measurements that are absent in continuous time. The necessary and sufficient reconstructability conditions established in this paper explain why these discretization-induced zeros can degrade disturbance reconstruction accuracy.

\textbf{6) Unified Geometric Interpretation.} We provide a unified geometric perspective for explaining disturbance reconstruction across the ASO family and clarifying its connection to UIO theory. This perspective reveals the common geometric mechanisms underlying the two observer paradigms, as well as their distinctions.
}

The rest of the paper is organized as follows. Section \ref{sec:preliminary} presents the problem formulation and preliminaries. Section \ref{sec:MBESO} details the two proposed observer designs. Section \ref{sec:errorCharacteristics} discusses the theoretical properties of the observers. Simulation results are provided in Section \ref{sec:simulation}. Section \ref{sec:conclusion} concludes the paper.

\textit{Notation:} \(\mathbb{R}\) and \(\mathbb{C}\) denote the set of real and complex numbers, respectively. \(\mathbb{R}^n\) represents the \(n\)-dimensional Euclidean space. The capital letter \(A\) represents the matrix. The lowercase letters \(c^0_i\) and \(e^0_i\) denote the \(i\)-th row and the \(i\)-th column of the output system matrix \(C_0\) and the input system matrix \(E_0\), respectively. \([\cdot]^{\top}\) denotes the transpose of \([\cdot]\). \([\cdot]_{p\times n}\) stands for a matrix \([\cdot]\) with dimensions of \(p\) by \(n\). \(I_p\) denotes an identity matrix of \(p\) dimensions. \(\check{A}_0=[\check{A}^0_{ij}]\), \(i,j=1,\cdots,p\) denotes that \(\check{A}_0\) is made up of some submatrices \(\check{A}^0_{ij}\) located in the \(i\)-th row and \(j\)-th column. \(\tilde{f}(z)\) denotes the \(z\)-transform of a function \(f(k)\). 

\section{Problem Formulation and Preliminaries}
\label{sec:preliminary}
Consider the following MIMO linear time-invariant 
discrete-time system with uncertainty
\begin{equation}\label{eq_1}
    \begin{cases}
        x(k+1)=A_0x(k)+B_0u(k)+E_0 f(k)\\
        y(k)=C_0x(k)
    \end{cases}
\end{equation}
where \(x\in\mathbb{R}^n\), \(u\in\mathbb{R}^m\), \(f\in\mathbb{R}^p\), and \(y\in\mathbb{R}^p\) represent the state vector, the known inputs, the disturbances including model uncertainties and external disturbances, and the outputs, respectively, and \(A_0\), \(B_0\), \(E_0\) and \(C_0\) are real and known matrices with appropriate dimensions. Without loss of generality, we assume that \(C_0\) has a full row rank. The dimension of \(f\) is assumed to be \(p\), {based on the well-known fact that the maximum number of independent disturbances that can be decoupled in observer design is determined by the number of available measurements in \(y\) \cite{willems1981disturbance}. The square case is adopted primarily for notational convenience and does not compromise the capability of the framework, since the two non-square cases can be handled straightforwardly. In over-measured scenarios, where the number of sensors exceeds the number of independent disturbance paths, measurement redundancy allows some measurements to be omitted without introducing additional dimensions into the observer. In under-measured scenarios, where the system has more independent disturbances than available measurements, prior work has demonstrated that disturbance decoupling is not guaranteed \cite{willems1981disturbance}.} The following assumption is adopted throughout the paper.

\begin{assumption}\label{assum1}
    For system \eqref{eq_1}, we have
    \(c_i^0 A_0^\kappa e_j^0=0, ~\forall 1\leq i,j\leq p\) and \(\forall \kappa<r_i-1\), and \(r_1+\cdots+r_p=n\). The \(p\times p\) disturbance gain matrix
        \begin{equation}\label{eq:V_0}
            V_0\triangleq\begin{bmatrix}
                c_1^0 A_0^{r_1-1}e_1^0 & \cdots & c_1^0 A_0^{r_1-1}e_p^0\\ 
                c_2^0 A_0^{r_2-1}e_1^0 & \cdots & c_2^0 A_0^{r_2-1}e_p^0\\
                \vdots & \ddots & \vdots\\
                c_p^0 A_0^{r_p-1}e_1^0 & \cdots & c_p^0 A_0^{r_p-1}e_p^0
            \end{bmatrix}
        \end{equation}
        is nonsingular. 
\end{assumption}

Assumption \ref{assum1} implies that system \eqref{eq_1} has a well-defined disturbance vector relative degree \(\{r_1, \cdots, r_p\}\) and no disturbance zero dynamics. For a system without a well-defined disturbance vector relative degree, a dynamic extension {has bee employed in \cite{wu2020performance} to established one for a multivariable invertible nonlinear system.} Without loss of generality, the known control input \(u\) is omitted from the subsequent analysis, {a standard simplification widely adopted within the UIO literature \cite{sundaram2007delayed}.} 


The primary objective is to {establish the necessary and sufficient conditions for the existence of a linear augmented observer to reconstruct the disturbances \(f\) from the outputs \(y\), while fully characterizing the transient performance of the disturbance reconstruction.}

Using Assumption \ref{assum1}, the following table can be constructed
\begin{equation}\label{eq:observabilityMatrix}
    \begin{tabular}{cccc}
        \(c_1^0\) & \(c_2^0\) & \multirow{4}*{\(\cdots\)} & \(c_p^0\) \\
        \(c_1^0 A_0\) & \(c_2^0A_0\) & \quad & \(c_p^0 A_0\) \\
        \(\vdots\) & \(\vdots\) & \quad & \(\vdots\) \\
        \(c_1^0 A_0^{r_1-1}\) & \(c_2^0A_0^{r_2-1}\) & \quad & \(c_p^0A_0^{r_p-1}\)
    \end{tabular}.
\end{equation}

\begin{lemma}\label{lem:observability}
    Under Assumption \ref{assum1}, system \eqref{eq_1} is observable. 
\end{lemma}

\begin{pf}
    From \eqref{eq:V_0}, we have \(V_0=[(c_1^0 A_0^{r_1-1})^{\top}, \cdots, (c_p^0A_0^{r_p-1})^{\top}]^{\top} E_0\). Since \(\text{rank}V_0=p\) and \(\text{rank}V_0\leq\min\{\text{rank}[(c_1^0 A_0^{r_1-1})^{\top},\) \(\cdots, (c_p^0A_0^{r_p-1})^{\top}]^{\top},\text{rank}E_0\}\), it follows that \(c_1^0 A_0^{r_1-1},\cdots,c_p^0A_0^{r_p-1}\) are linearly independent. From the dual of Property 3.3.4 in \cite[p. 150]{basile1992controlled} and its logically equivalent contrapositive, we have that the matrix constructed by \eqref{eq:observabilityMatrix} is full rank. From Assumption \ref{assum1}, we also have \(r_1+\cdots+r_p=n\). Then, system \eqref{eq_1} is observable. \hfill $\square$
\end{pf}

According to Lemma \ref{lem:observability}, the matrix constructed by \eqref{eq:observabilityMatrix} constitutes an observability matrix. The constants \(r_i\) for \(i=1, \cdots, p\) are the output structural indices (also known as the observability indices), and \(r_{\max}=\max(r_1, \cdots, r_p)\) denotes the observability index. Based on the construction of the observability matrix, the order of \(c_1^0, \cdots, c_p^0\) can be rearranged such that each column ends with a vector that is linearly independent of all previous entries. Therefore, the vector \(c_i^0 A_0^{r_i}\) can be expressed as a linear combination of all previous entries:
\begin{equation}\label{eq:canonicalLastLine}
    c_i^0A_0^{r_i}=\sum_{j=1}^p\sum_{\kappa=0}^{r_i-1}a_{ij\kappa}c_j^0 A_0^{\kappa}+\sum_{j<i}a_{ijr_i}c_j^0A_0^{r_i}, ~ i=1, \cdots, p,
\end{equation}
where the coefficient \(a_{ij\kappa}\) is zero for \(\kappa\geq r_j\). 

To facilitate the proof of the theorems in Section \ref{sec:errorCharacteristics}, the observability canonical form is transformed into the observer companion form in this paper, cf. \cite[p. 150]{basile1992controlled}. \eqref{eq:canonicalLastLine} can be written as 
\begin{equation}\label{eq:canonicalLastLineVariant}
    (c_i^0-\sum_{j<i}a_{ijr_i}c_j^0)A_0^{r_i}=\sum_{j=1}^p\sum_{\kappa=0}^{r_i-1}a_{ij\kappa}c_j^0 A_0^{\kappa}, ~ i=1, \cdots, p.
\end{equation}
The newly assumed output matrix can be defined as 
\begin{equation}\label{eq:newMeasurement}
    \bar{\bar{c}}_i^0\triangleq c_i^0-\sum_{j<i}a_{ijr_i}c_j^0, ~ i=1, \cdots, p. 
\end{equation} 
To facilitate analysis in subsequent sections, \eqref{eq:newMeasurement} is expressed in a matrix form as \(\bar{\bar{C}}_0=PC_0\), where
\begin{equation}\label{eq:newMeasurementMatrixForm}
    \bar{\bar{C}}_0=\begin{bmatrix}
        \bar{\bar{c}}_1^0\\
        \bar{\bar{c}}_2^0\\
        \vdots\\
        \bar{\bar{c}}_p^0
    \end{bmatrix}, P=\begin{bmatrix}
        1 & 0 & \cdots & 0\\
        -a_{21r_2} & 1 & \cdots & 0\\
        \vdots & \vdots & \ddots & \vdots\\
        -a_{p1r_p} & -a_{p2r_p} & \cdots & 1
    \end{bmatrix}. 
\end{equation}
Then, \eqref{eq:canonicalLastLineVariant} is in the form of 
\begin{equation}\label{eq:bParameters}
    \bar{\bar{c}}_i^0 A_0^{r_i}=\sum_{j=1}^p\sum_{\kappa=0}^{r_i-1}b_{ij\kappa}\bar{\bar{c}}_j^0 A_0^{\kappa}, ~ i=1, \cdots, p,
\end{equation}
where the coefficient \(b_{ij\kappa}\) is zero for \(\kappa\geq r_j\). The corresponding observability matrix, which is similar to \eqref{eq:observabilityMatrix}, can be constructed from the following table:
\begin{equation}\label{eq:observabilityMatrixNew}
    \begin{tabular}{cccc}
        \(\bar{\bar{c}}_1^0\) & \(\bar{\bar{c}}_2^0\) & \multirow{4}*{\(\cdots\)} & \(\bar{\bar{c}}_p^0\) \\
        \(\bar{\bar{c}}_1^0 A_0\) & \(\bar{\bar{c}}_2^0A_0\) & \quad & \(\bar{\bar{c}}_p^0 A_0\) \\
        \(\vdots\) & \(\vdots\) & \quad & \(\vdots\) \\
        \(\bar{\bar{c}}_1^0 A_0^{r_1-1}\) & \(\bar{\bar{c}}_2^0A_0^{r_2-1}\) & \quad & \(\bar{\bar{c}}_p^0A_0^{r_p-1}\)
    \end{tabular}.
\end{equation}

Although the original output matrix \(C_0\) is replaced by the newly assumed output matrix \(\bar{\bar{C}}_0\) in \eqref{eq:newMeasurementMatrixForm}, we have the following lemma. 

\begin{lemma}\label{lem:V_0Invariant}
    If Assumption \ref{assum1} holds, the disturbance gain matrix \(V_0\) in \eqref{eq:V_0} remains unchanged under the new output matrix assumed in \eqref{eq:newMeasurement}, i.e., \(V_0=\bar{\bar{V}}_0\), where
    \begin{equation}\label{eq:V_0New}
        \bar{\bar{V}}_0\triangleq\begin{bmatrix}
            \bar{\bar{c}}_1^0 A_0^{r_1-1}e_1^0 & \cdots & \bar{\bar{c}}_1^0 A_0^{r_1-1}e_p^0\\ 
            \bar{\bar{c}}_2^0 A_0^{r_2-1}e_1^0 & \cdots & \bar{\bar{c}}_2^0 A_0^{r_2-1}e_p^0\\
            \vdots & \ddots & \vdots\\
            \bar{\bar{c}}_p^0 A_0^{r_p-1}e_1^0 & \cdots & \bar{\bar{c}}_p^0 A_0^{r_p-1}e_p^0
        \end{bmatrix}.
    \end{equation}
\end{lemma}

\begin{pf}
    The \(i\)-th row of the right matrix in \eqref{eq:V_0New} is 
    \begin{equation}\label{eq:V_0NewProof}
        \bar{\bar{c}}_i^0 A_0^{r_i-1}E_0= c_i^0 A_0^{r_i-1}E_0 -\sum_{j<i}a_{ijr_i}c_j^0 A_0^{r_i-1}E_0,
    \end{equation}
    where the coefficient \(a_{ijr_{i}}\) is zero for \(r_{i}\geq r_j\). From Assumption \ref{assum1}, we have \(c_j^0 A_0^{\kappa}E_0=0_{1\times p}\) for \(1\leq j \leq p\) and \(\kappa<r_j-1\). Substituting \(\kappa=r_i-1\) into the above equation yields \(c_j^0 A_0^{r_i-1}E_0=0_{1\times p}\) for \(r_i<r_j\). Therefore, in \eqref{eq:V_0NewProof}, \(\sum_{j<i}a_{ijr_i}c_j^0 A_0^{r_i-1}E_0=0\). \hfill $\square$
\end{pf}

\section{Model-Based Extended State Observer Design}
\label{sec:MBESO}
\begin{figure}[htbp]
    \begin{center}
        \includegraphics[width=3.5 in]{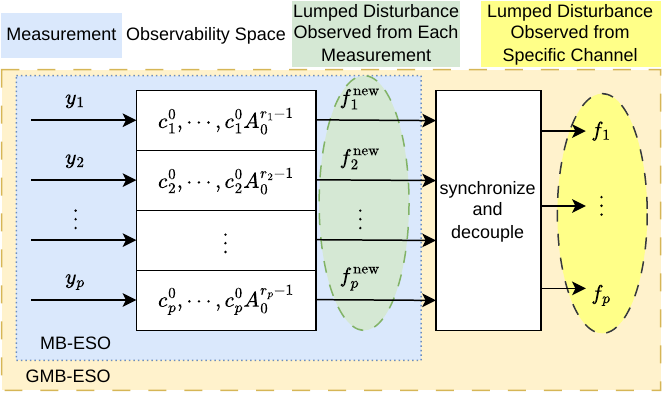}
        \caption{{Structural relationship and architectural comparison between the proposed MB-ESO and GMB-ESO.} }
        \label{fig:DisturbanceObserver}
    \end{center}
\end{figure}

In this section, two observer designs are proposed according to the structure of the disturbance gain matrix $V_0$ in \eqref{eq:V_0}. When $V_0$ is diagonal, the observer is referred to as the MB-ESO. Otherwise, a GMB-ESO is developed. {The structural relationship between these two observers is illustrated in Fig. \ref{fig:DisturbanceObserver}.  
Within this framework, each measurement channel reconstructs a distinct lumped disturbance. Consequently, the MB-ESO directly reconstruct these disturbances when the disturbance gain matrix is diagonal. For a non-diagonal gain matrix, however, cross-channel coupling causes the reconstructed disturbances to exhibit different propagation delays due to different relative degrees. The GMB-ESO resolves this issue through an algebraic synchronization mechanism that aligns the delayed channels, enabling safe inversion of the disturbance gain matrix to recover the true physical lumped disturbances.}


\subsection{A Model-Based Extended State Observer}
In ESO design, the disturbance vector \(f\) is regarded as an extended state, and an observer is designed to reconstruct the resulting augmented state. 
The augmented system can be written as
\begin{equation}\label{eq_2}
    \begin{cases}
        \underline{x}(k+1) = A \underline{x}(k) + Bu(k) + E \Delta f(k) \\
        y(k) = C \underline{x}(k)
    \end{cases}
\end{equation}
where \(\underline{x}(k)=\begin{bmatrix}
    x(k)\\
    f(k)
\end{bmatrix}\), 
\(
    A = \begin{bmatrix}
    A_0 & E_0\\
    0_{p \times n} & I_{p}
    \end{bmatrix}
\), 
\(  
    B = \begin{bmatrix}
    B_0\\
    0_{p\times m}
    \end{bmatrix}
\), 
\(  E = [0_{p\times n}, I_p]^{\top}
\), 
\(  C=[C_0, 0_{p\times p}]
\), 
and \(\Delta f(k) = f(k+1) - f(k)\) represents the variation of the true disturbance. 

\begin{lemma}\label{lem:systemObservability}
    If Assumption \ref{assum1} holds, then the augmented system \eqref{eq_2} is observable. 
\end{lemma}

\begin{pf}
    Under Assumption \ref{assum1}, {along with Lemmas \ref{lem:observability} and \ref{lem:V_0Invariant}}, there exists an invertible matrix \(S_1\) such that \((A, E, C)\) is equivalent to \((A_1, E_1, C_1)\), \emph{i.e.}, 
    \begin{equation}\label{lem_eq1}
        A_1=S_1 A S_1^{-1}, E_1=S_1 E, C_1=C S_1^{-1},
    \end{equation}
    where 
    \(A_1=\begin{bmatrix}
        \bar{A}_0 & \acute{E}_0\\
        0_{p\times n} & I_p
    \end{bmatrix}\), \(E_1=\begin{bmatrix}
        0_{n\times p}\\
        V_0
    \end{bmatrix}\), 
    \(S_1=\begin{bmatrix}
        S_0 & 0_{n\times p}\\
        0_{p\times n} & V_0
    \end{bmatrix}\), 
    \(S_0=[(\bar{\bar{c}}_1^0)^{\top}, \cdots, (\bar{\bar{c}}_1^0A_0^{r_1-1})^{\top}, \cdots, (\bar{\bar{c}}_p^0)^{\top}, \cdots, (\bar{\bar{c}}_p^0A_0^{r_p-1})^{\top}]^{\top}\), \(\bar{A}_0=[\bar{A}^0_{ij}]\), \(i,j=1,\cdots,p\), 
    \(\bar{A}_{ij}^0=\begin{cases}
        \begin{bmatrix}\begin{smallmatrix}
            0 & 0 & \cdots & 0 & \cdots & 0\\
            \vdots & \vdots & \ddots & \vdots & \ddots & \vdots\\
            0 & 0 & \cdots & 0 & \cdots & 0\\
            b_{ij0} & b_{ij1} & \cdots & b_{ij(r_i-1)} & \cdots & 0
            \end{smallmatrix}
        \end{bmatrix}_{r_i\times r_j} & \text{if } r_i<r_j\\
        \begin{bmatrix}
            \begin{smallmatrix}
                0 & 0 & \cdots & 0\\
                \vdots & \vdots & \ddots & \vdots\\
                0 & 0 & \cdots & 0\\
                b_{ij0} & b_{ij1} & \cdots & b_{ij(r_j-1)}
            \end{smallmatrix}
        \end{bmatrix}_{r_i\times r_j} & \text{if } r_i\geq r_j
    \end{cases}\), 
    \(i\neq j\), \(\bar{A}_{ii}^0=\begin{bmatrix}
        0 & 1 & \cdots & 0\\
        \vdots & \vdots & \ddots & \vdots\\
        0 & 0 & \cdots & 1\\
        b_{ii0} & b_{ii1} & \cdots & b_{ii(r_i-1)}
    \end{bmatrix}_{r_i\times r_i}\), \(i=j\), \(C_1=[\bar{C}_0, 0_{p\times p}]\), \(\bar{C}_0=[\bar{C}^0_1, \cdots, \bar{C}^0_p]_{p\times n}\), \(\bar{C}_i^0=[\times_i, 0_{p\times (r_i-1)}]_{p\times r_i}\), and the \(\times_i\) is the \(i\)-th column of the inverse of \(P\) in \eqref{eq:newMeasurementMatrixForm}, 
    \[\acute{E}_0=\begin{bmatrix}
        \acute{E}_1^0\\
        \vdots\\
        \acute{E}_p^0
    \end{bmatrix}_{n\times p}, \quad 
    \acute{E}_i^0=\begin{bmatrix}
        0 & \cdots & 0 & \cdots & 0\\
        \vdots & \ddots & \vdots & \ddots & \vdots\\
        0 & \cdots & 1 & \cdots & 0
    \end{bmatrix}_{r_i \times p},\] 
    the \(1\) in the last row of \(\acute{E}_i^0\) occurs at the \(i\)-th column, the other entries in \(\acute{E}_i^0\) are zero, \(i=1,\cdots,p\). 

    The system \((A_1, E_1, PC_1)\) {represents} a combination of observability canonical forms, {matching the structure utilized} in conventional ESO, {where} \(P\) is nonsingular. {Then,} each subsystem of \((A_1, PC_1)\) is observable. Since the observability property is invariant under any equivalent transformation, \((A, C)\) is also observable. \hfill $\square$
\end{pf}

According to Lemma \ref{lem:systemObservability}, a MB-ESO for system \eqref{eq_2} can be constructed as follows: 
\begin{equation}\label{eq_3}
    \hat{\underline{x}}(k+1) = A\hat{\underline{x}}(k) + Bu(k) + L(y(k)-C\hat{\underline{x}}(k)),
\end{equation}
where \(\hat{\underline{x}}=[\hat{x}^{\top}, \hat{f}^{\top}]^{\top}\) is a reconstruction of the state \(\underline{x}\), and \(L \in \mathbb{R} ^{(n+p)\times p}\) is the observer gain matrix to be determined in this paper. From Lemma \ref{lem:systemObservability}, the eigenvalues of \(A-LC\) can be arbitrarily placed inside the unit circle. For the sake of simplicity, all the eigenvalues of \(A-LC\) for each subsystem are placed at \(\omega_{o_i}\), \(i=1,\cdots,p\). Note that, for continuous-time systems, the eigenvalues are all placed at \(-\omega_{o_i}\), where \(\omega_{o_i}\) is denoted as the observer bandwidth of ESO \cite{gao2003scaling}. Unlike other ESOs, the Luenberger state observer~\eqref{eq_3} is referred to here as MB-ESO, since the matrices $A_0$, $B_0$, $C_0$, and $E_0$ are in general form with model information and multiple measurements. 

Based on \eqref{eq_2} and \eqref{eq_3}, the error dynamics of 
MB-ESO \eqref{eq_3} is given by: 
\begin{equation}\label{eq_5}
    e(k+1) = (A-LC)e(k) + E\Delta f(k),
\end{equation}
where \(e(k) = \underline{x}(k) - \hat{\underline{x}}(k)\), and \(E\Delta f(k)\) results from the neglect of the unknown input \(\Delta f\) in \eqref{eq_3}.

\subsection{A Generalized Model-Based Extended State Observer}
The MB-ESO in \eqref{eq_3} can accurately reconstruct \(f\) only when \(V_0\) is diagonal, as will be proved in Theorem \ref{thm2}. However, this condition may not always hold in practice. Hence, the MB-ESO in \eqref{eq_3} is modified accordingly.

The system \eqref{eq_1} can be rewritten as 
\begin{equation}\label{eq:systemRewritten}
    \begin{cases}
        x(k+1)=A_0x(k)+B_0u(k)+E_0 V_0^{-1} f_{\text{new}}(k)\\
        y(k)=C_0x(k)
    \end{cases}
\end{equation}
where \(f_{\text{new}}(k)=V_0 f(k)\). Then, the \(V_{0}^{\text{new}}\) calculated by \eqref{eq:V_0} for the revised system \eqref{eq:systemRewritten} is an identity matrix, which is diagonal. The newly augmented system can be written as
\begin{equation}\label{eq:newAugmentedSystem}
    \begin{cases}
        \underline{x}(k+1) = A \underline{x}(k) + Bu(k) + E \Delta f_{\text{new}}(k) \\
        y(k) = C \underline{x}(k)
    \end{cases}
\end{equation}
where \(\underline{x}(k)=\begin{bmatrix}
    x(k)\\
    f_{\text{new}}(k)
\end{bmatrix}\), 
\(
    A = \begin{bmatrix}
    A_0 & E_0V_0^{-1}\\
    0_{p \times n} & I_{p}
    \end{bmatrix}
\), 
\(  
    B = \begin{bmatrix}
    B_0\\
    0_{p\times m}
    \end{bmatrix}
\), 
\(  E = \begin{bmatrix}
    0_{n\times p}\\
    I_p
\end{bmatrix}
\), 
\(  C=[C_0, 0_{p\times p}]
\), 
and \(\Delta f_{\text{new}}(k) = f_{\text{new}}(k+1) - f_{\text{new}}(k)\). 

Since \(V_{0}^{\text{new}}\) is diagonal, a GMB-ESO for system \eqref{eq:systemRewritten} can be designed as
\begin{equation}\label{eq:newMBESO}
    \hat{\underline{x}}(k+1)=A\hat{\underline{x}}(k)+Bu(k)+L(y(k)-C\hat{\underline{x}}(k)),
\end{equation}
where \(\hat{\underline{x}}=[\hat{x}^{\top},\hat{f}_{\text{new}}^{\top}]^{\top}\) is a reconstruction of the state \(\underline{x}\). From Theorem \ref{thm2}, we have that \(\hat{f}_i^{\text{new}}(k)\) is a reconstruction of \(f_i^{\text{new}}(k)\) with different delay \(r_i+1\). This delay is related to the output structural index of the corresponding output. To obtain the reconstruction of the original \(f\), {the individual channels \(\hat{f}^{\text{new}}_i(k)\) must be synchronized to achieve an identical delay}
\begin{equation}\label{eq:delayF}
    \hat{\bar{f}}_{\text{new}}(k)=[\hat{f}^{\text{new}}_1(k-(r_{\max}-r_1)),\cdots,\hat{f}^{\text{new}}_p(k-(r_{\max}-r_p))]^{\top}.
\end{equation}
Then, the reconstruction of the original \(f\) is 
\begin{equation}\label{eq:newf}
    \hat{f}(k)=V_0^{-1}\hat{\bar{f}}_{\text{new}}(k). 
\end{equation}
For general multivariable systems, the GMB-ESO for system \eqref{eq:systemRewritten} is given by \eqref{eq:newMBESO}--\eqref{eq:newf}.

\section{Analysis of Disturbance Estimation for MB-ESO and GMB-ESO}
\label{sec:errorCharacteristics}
In this section, the disturbance estimation performance of the MB-ESO and GMB-ESO is analyzed. 
Specifically, finite-step convergence of the disturbance reconstruction is first established. 
Then, the transient behaviors of the disturbance estimation error, including error bounds and convergence behavior, are investigated.

\subsection{Finite-Step Convergence}
The error dynamics of MB-ESO in \eqref{eq_5} contains an unknown true disturbance's variation \(\Delta f\). The reconstructed disturbance can still converge to the true disturbance with a time delay, as shown in the following theorems. {The core strategy for proving Theorem 1 relies on synthesizing an observer gain matrix \(L\) capable of enforcing complete subsystem decoupling while simultaneously assigning the observer's eigenvalues. For the detailed construction of the similarity transformation matrices utilized throughout this proof, the reader is kindly referred to \cite[p. 150]{basile1992controlled}. To facilitate a thorough understanding of the proof, the complete MATLAB source code implementing a numerical example, both with and without the output transformation matrix \(P\), is publicly available on Anonymous GitHub\footnote{https://anonymous.4open.science/r/MB-ESO-B824/README.md}.} 

\begin{theorem}\label{thm2}
    Condition \eqref{eq_thm1} holds for observer \eqref{eq_3}: 
    \begin{equation}\label{eq_thm1}
        \hat{f}_i(k)=f_i(k-r_i-1), ~~\forall i=1,\cdots,p ~\text{and} ~k>r_i+1, 
    \end{equation}
    if and only if a) system \eqref{eq_1} satisfies Assumption \ref{assum1}, b) \(V_0\) is diagonal, c) all eigenvalues of the observer are placed at the origin, and d) all subsystems of the observer are decoupled. 
\end{theorem}

\begin{pf}
    The estimation error dynamics of observer \eqref{eq_3} for augmented system \eqref{eq_2} is \eqref{eq_5}. Then, by using Lemma \ref{lem:systemObservability}, \eqref{eq_5} can be written as
    \begin{equation}\label{eq_8}
        e_1(k+1)=(A_1-L_1C_1)e_1(k)+E_1 \Delta f(k),
    \end{equation}
    where \(e_1(k)=S_1e(k)\), \(L_1=S_1L\). \((A_1, C_1)\) can be converted to a pseudo-observability canonical form with an invertible matrix \(S_2\) with 
    \begin{equation}\label{eq_9}
        A_2=S_2A_1S_2^{-1}, ~C_2=C_1S_2^{-1},
    \end{equation}
    where {\(S_2=[(\ddot{c}^1_1)^{\top},\cdots,(\ddot{c}^1_1A_1^{r_1})^{\top},\cdots,(\ddot{c}^1_p)^{\top},\cdots,(\ddot{c}^1_pA_1^{r_p})^{\top}]^{\top}\), \(\ddot{c}^1_i\) is the \(i\)-th row of \(\ddot{C}_1\), \(\ddot{C}_1=[\dot{C}_0, 0_{p\times p}]\), \(\dot{C}_0=[\dot{C}^0_1, \cdots, \dot{C}^0_p]_{p\times n}\), \(\dot{C}_j^0=[\mathfrak{X}_j, 0_{p\times (r_j-1)}]_{p\times r_j}\), and the column vector \(\mathfrak{X}_j\) contains a single \(1\) at the \(j\)th row, the all other entries in \(\mathfrak{X}_j\) are zero,} 
    \(C_2=[C^2_1, C^2_2, \cdots, C^2_p]_{p\times (n+p)}\), \(C^2_i=[\times_i,0_{p\times r_i}]_{p\times{(r_i+1)}}\), the \(\times_i\) is the \(i\)-th column of \(P^{-1}\), \(A_2=[A^2_{ij}]\), \(i,j=1,\cdots,p\), 
    \(A^2_{ii}=\begin{bmatrix}
        \begin{smallmatrix}
        0 & 1 & \cdots & 0 & 0\\
        0 & 0 & \cdots & 0 & 0\\
        \vdots & \vdots & \ddots & \vdots & \vdots \\
        0 & 0 & \cdots & 0 & 1\\
        -b_{ii0} & b_{ii0}-b_{ii1} & \cdots & b_{ii(r_i-2)}-b_{ii(r_i-1)} & b_{ii(r_i-1)}+1
        \end{smallmatrix}
    \end{bmatrix}\), \(i=j\), 
    \(A^2_{ij}=\begin{cases}
        \begin{bmatrix}
            \begin{smallmatrix}
                0 & 0 & \cdots & 0 & \cdots & 0\\
                \vdots & \vdots & \ddots & \vdots & \ddots & \vdots\\
                0 & 0 & \cdots & 0 & \cdots & 0\\
                -b_{ij0} & b_{ij0}-b_{ij1} & \cdots & b_{ij(r_i-1)} & \cdots & 0
            \end{smallmatrix}
        \end{bmatrix} & \text{if } r_i<r_j\\
        \begin{bmatrix}
            \begin{smallmatrix}
                0 & 0 & \cdots & 0\\
                \vdots & \vdots & \ddots & \vdots\\
                0 & 0 & \cdots & 0\\
                -b_{ij0} & b_{ij0}-b_{ij1} & \cdots & b_{ij(r_j-1)}
            \end{smallmatrix}
        \end{bmatrix} & \text{if } r_i\geq r_j
    \end{cases}\), \(i\neq j\), and the dimensions of \(A^2_{ii}\) and \(A^2_{ij}\) are \((r_i+1)\times (r_i+1)\) and \((r_i+1)\times (r_j+1)\), respectively. 

    Thus, the estimation error in \eqref{eq_8} can be transformed into
    \begin{equation}\label{eq_10}
        e_2(k+1)=(A_2-L_2C_2)e_2(k)+E_2\Delta f(k),
    \end{equation}
    where \(e_2(k)=S_2e_1(k)\), \(L_2=S_2L_1\), \(E_2=S_2E_1=[(E^2_1)^{\top},(E^2_2)^{\top},\cdots,(E^2_p)^{\top}]^{\top}_{p\times(n+p)}\), \(E^2_i=\begin{bmatrix}
        0_{r_i\times p}\\
        v^0_i
    \end{bmatrix}_{(r_i+1)\times p}\), \(i=1,\cdots, p\), 
    and \(v_i^0\) is the \(i\)-th row of \(V_0\).  

    Note that the matrix \(A_2\) can be transformed into the following pseudo-observer companion form \(A_3\) using an invertible matrix \(Q_1\), where \(A_3=[A^3_{ij}]\), \(i,j=1,\cdots,p\), 
    \(A^3_{ii}=\begin{bmatrix}
        b_{ii(r_i-1)}+1 & 1 & \cdots & 0 & 0\\
        b_{ii(r_i-2)}-b_{ii(r_i-1)} & 0 & \cdots & 0 & 0\\
        \vdots & \vdots & \ddots & \vdots & \vdots\\
        b_{ii0}-b_{ii1} & 0 & \cdots & 0 & 1\\
        -b_{ii0} & 0 & \cdots & 0 & 0
    \end{bmatrix}_{(r_i+1)\times(r_i+1)}\), \(i= j\),
    \(A^3_{ij}=\begin{cases}
        \begin{bmatrix}\begin{smallmatrix}
            b_{ij(r_i-1)} & 0 & \cdots & 0\\
            b_{ij(r_i-2)}-b_{ij(r_i-1)} & 0 & \cdots & 0\\
            \vdots & \vdots & \ddots & \vdots\\
            b_{ij0}-b_{ij1} & 0 & \cdots & 0\\
            -b_{ij0} & 0 & \cdots & 0
            \end{smallmatrix}
        \end{bmatrix}_{(r_i+1)\times (r_j+1)} & \text{if }r_i< r_j\\
        \begin{bmatrix}\begin{smallmatrix}
            0 & 0 & \cdots & 0\\
            \vdots & \vdots & \ddots & \vdots\\
            b_{ij(r_j-1)} & 0 & \cdots & 0\\
            \vdots & \vdots & \ddots & \vdots\\
            b_{ij0}-b_{ij1} & 0 & \cdots & 0\\
            -b_{ij0} & 0 & \cdots & 0
            \end{smallmatrix}
        \end{bmatrix}_{(r_i+1)\times (r_j+1)} & \text{if } r_i\geq r_j
    \end{cases}\), \(i\neq j\), and \(Q_1=[Q^1_{ij}]\), \(i,j=1,\cdots,p\), 
    \[Q^1_{ii}=\begin{bmatrix}\begin{smallmatrix}
        1 & 0 & \cdots & 0 & 0\\
        -1-b_{ii(r_i-1)} & 1 & \cdots & 0 & 0\\
        \vdots & \vdots & \ddots & \vdots & \vdots\\
        b_{ii2}-b_{ii1} & b_{ii3}-b_{ii2} & \cdots & 1 & 0\\
        b_{ii1}-b_{ii0} & b_{ii2}-b_{ii1} & \cdots & -1-b_{ii(r_i-1)} & 1
        \end{smallmatrix}
    \end{bmatrix}, i=j,\]
    \[Q^1_{ij}=\begin{cases}
        \begin{bmatrix}\begin{smallmatrix}
            0 & 0 & \cdots & 0 & \cdots & 0\\
            -b_{ij(r_i-1)} & 0 & \cdots & 0 & \cdots & 0\\
            \vdots & \vdots & \ddots & \vdots & \ddots & \vdots\\
            b_{ij2}-b_{ij1} & b_{ij3}-b_{ij2} & \cdots & 0 & \cdots & 0\\
            b_{ij1}-b_{ij0} & b_{ij2}-b_{ij1} & \cdots & -b_{ij(r_i-1)} & \cdots & 0
            \end{smallmatrix}
        \end{bmatrix} & \text{if } r_i<r_j\\
        \begin{bmatrix}
            \begin{smallmatrix}
                0 & 0 & \cdots & 0 & 0\\
                \vdots & \vdots & \ddots & \vdots & \vdots\\
                -b_{ij(r_j-1)} & 0 & \cdots & 0 & 0\\
                \vdots & \vdots & \ddots & \vdots & \vdots\\
                b_{ij2}-b_{ij1} & b_{ij3}-b_{ij2} & \cdots & 0 & 0\\
                b_{ij1}-b_{ij0} & b_{ij2}-b_{ij1} & \cdots & -b_{ij(r_j-1)} & 0
            \end{smallmatrix}
        \end{bmatrix} & \text{if }r_i\geq r_j
    \end{cases},\]
    \(i\neq j\), and the dimensions of \(Q^1_{ii}\) and \(Q^1_{ij}\) are \((r_i+1)\times (r_i+1)\) and \((r_i+1)\times (r_j+1)\), respectively. {Consequently, decoupling the subsystems and achieving the desired eigenvalue assignment can be accomplished by strictly modifying the columns of \(A_{3}\) that correspond directly to the output channels.}
    
    By using the invertible matrix \(Q_1\), estimation error dynamics in \eqref{eq_10} can be written as
    \begin{equation}\label{eq_11}
        e_3(k+1)=(A_3-L_3C_3)e_3(k)+E_3\Delta f(k),
    \end{equation}
    where \(e_3(k)=Q_1e_2(k)\), \(A_3=Q_1A_2Q_1^{-1}\), \(L_3=Q_1L_2\), \(C_3=C_2Q_1^{-1}=C_2\), and \(E_3=Q_1E_2=E_2\). {The observer gain matrix \(L_{3}\) is synthesized by decoupling the constituent subsystems of \(A_3 - L_3 C_3\) and placing their corresponding eigenvalues. Determining \(L_{3}\) requires first calculating \(\bar{L}_{3}\) in the following equation}
    \begin{equation}\label{eq:L3_bar}
        A_3-L_3C_3=A_3-L_3P^{-1}\bar{C}_{3}=A_3-\bar{L}_{3}\bar{C}_{3},
    \end{equation}
    where \(\bar{C}_3=[\bar{C}^3_1,\cdots,\bar{C}^3_p]_{p\times (n+p)}\), \(\bar{C}^3_i=[\mathfrak{X}_i,0_{p\times r_i}]_{p\times (r_i+1)}\), the column vector \(\mathfrak{X}_i\) contains a single \(1\) in the \(i\)-th row, and all other entries in \(\mathfrak{X}_i\) are zero, \(i=1,\cdots,p\). 
    
    Let \(\bar{L}_3=[\bar{\underline{l}}^3_{ij}]\), \(i,j=1,\cdots,p\), and \(\bar{\underline{l}}^3_{ij}=[\bar{l}^{ij}_1,\cdots,\bar{l}^{ij}_{r_i+1}]^{\top}\). To decouple these \(p\) subsystems, \(\bar{\underline{l}}^3_{ij}, i\neq j\) can be chosen such that all off-diagonal submatrices of \(A_3-\bar{L}_3\bar{C}_3\) are zero matrices. For those submatrices on the diagonal, we have 
    \begin{equation}\label{eq:L3_bar_zero}
        A^3_{ii}-\bar{\underline{l}}_{ii}^3\bar{\underline{c}}_{ii}^3=\begin{bmatrix}\begin{smallmatrix}
            b_{ii(r_i-1)}+1-\bar{l}^{ii}_1 & 1 & \cdots & 0 & 0\\
            b_{ii(r_i-2)}-b_{ii(r_i-1)}-\bar{l}^{ii}_2 & 0 & \cdots & 0 & 0\\
            \vdots & \vdots & \ddots & \vdots & \vdots \\
            b_{ii0}-b_{ii1}-\bar{l}^{ii}_{r_i} & 0 & \cdots & 0 & 1\\
            -b_{ii0}-\bar{l}^{ii}_{r_i+1} & 0 & \cdots & 0 & 0
            \end{smallmatrix}
        \end{bmatrix}
    \end{equation}
    where \(\bar{\underline{c}}_{ii}^3\) is the \(i\)-th row of \(\bar{C}_{i}^3\).

    Moreover, the characteristic polynomial of the matrix 
    \(\begin{bmatrix}\begin{smallmatrix}
        -\alpha^{ii}_1 & 1 & 0 & \cdots & 0\\
        -\alpha^{ii}_2 & 0 & 1 & \cdots & 0\\
        \vdots & \vdots & \vdots & \ddots & \vdots\\
        -\alpha^{ii}_{r_i} & 0 & 0 & \cdots & 1\\
        -\alpha^{ii}_{r_i+1} & 0 & 0 & \cdots & 0
        \end{smallmatrix}
    \end{bmatrix}\) is \(z^{r_i+1}+\alpha^{ii}_1 z^{r_i}+\cdots+\alpha^{ii}_{r_i}z+\alpha^{ii}_{r_i+1}\). 
    Therefore, if all eigenvalues are at the origin, the characteristic polynomial of each subsystem is \(z^{r_i+1}\), which implies \(\alpha^{ii}_1=\alpha^{ii}_2=\cdots=\alpha^{ii}_{r_i+1}=0\). Then, by equating coefficients, we have \(\grave{A}_3=[\grave{A}^3_{ij}]\triangleq A_3-\bar{L}_3 \bar{C}_3\), \(i,j=1, \cdots, p\), where
    \[\grave{A}^3_{ij}=\begin{cases}
        \begin{bmatrix}
        0 & 1 & \cdots & 0 & 0\\
        \vdots & \vdots & \ddots & \vdots & \vdots \\
        0 & 0 & \cdots & 1 & 0\\
        0 & 0 & \cdots & 0 & 1\\
        0 & 0 & \cdots & 0 & 0
    \end{bmatrix}_{(r_i+1)\times (r_i+1)} & \text{if } i=j,\\
    0_{(r_i+1)\times (r_j+1)} & \text{if } i\neq j.
    \end{cases}\] 

    {After achieving subsystem decoupling and eigenvalue assignment, \eqref{eq_11} can be expressed as}
    \begin{equation}\label{eq_12}
        e_3(k+1)=\grave{A}_3e_3(k)+E_3 \Delta f(k). 
    \end{equation}
    {Taking the \(z\)-transform of \eqref{eq_12} while neglecting initial conditions leads to}
    \begin{equation}\label{eq_13}
        \tilde{e}_3(z)=(zI-\grave{A}_3)^{-1}E_3 \Delta \tilde{f}(z). 
    \end{equation}
    Since all subsystems are decoupled and all eigenvalues are placed at the origin, the off-diagonal submatrices of \((zI-\grave{A}_3)^{-1}\) are all zero matrices, and those matrices on the diagonal are 
    \begin{equation}\label{eq_14}
        (zI-\grave{A}_3)^{-1}_{ii}=\begin{bmatrix}
            1/z & 1/z^2 & \cdots & 1/z^{r_i+1}\\
            \vdots & \vdots & \ddots & \vdots\\
            0 & 0 & \cdots & 1/z^2\\
            0 & 0 & \cdots & 1/z
        \end{bmatrix}, ~i=1,\cdots,p.
    \end{equation}

    Next, substituting \(e_1 (k)=S_1 e(k)\), \(e_2 (k)=S_2 e_1 (k)\), and \(e_3 (k)=Q_1 e_2 (k)\) into \eqref{eq_13} yields
    \begin{equation}\label{eq_15}
        S_1\tilde{e}(z)=(Q_1S_2)^{-1}(zI-\grave{A}_3)^{-1}E_3\Delta \tilde{f}(z). 
    \end{equation}
    
    From \eqref{eq_10} and \eqref{eq_11}, {it follows that} 
    \begin{equation}\label{eq:newE}
        E_3=E_\text{new}V_0,
    \end{equation}
    where \(E_\text{new}=\begin{bmatrix}
        E^\text{new}_1\\
        \vdots\\
        E^\text{new}_p
    \end{bmatrix}\), \(E^\text{new}_i=\begin{bmatrix}\begin{smallmatrix}
        0 & \cdots & 0 & \cdots & 0\\
        \vdots & \ddots & \vdots & \ddots & \vdots\\
        0 & \cdots & 1 & \cdots & 0
        \end{smallmatrix}
    \end{bmatrix}_{(r_i+1)\times p}\), the \(1\) in the last row of \(E_i^\text{new}\) occurs at the \(i\)-th column, and the other entries in \(E_i^\text{new}\) are zero, \(i=1,\cdots,p\). 
    
    Substituting \eqref{eq_14} and \eqref{eq:newE} into \eqref{eq_15} yields 
    \begin{equation}\label{eq_16}
        S_1\tilde{e}(z)=(Q_1S_2)^{-1}\cdot\text{diag}(Z_1, \cdots, Z_p)\cdot V_0\Delta\tilde{f}(z),
    \end{equation}
    where \(Z_i=[1/z^{r_i+1}, 1/z^{r_i}, \cdots, 1/z]^{\top}\), \(i=1,\cdots,p\).

    Note that \(Q_1S_2=[[W_{ij}], E_{\text{new}}]\), \(i,j=1,\cdots,p\), where 
    \[W_{ii}=\begin{bmatrix}
            \begin{smallmatrix}
                1 & 0 & \cdots & 0\\
                -1-b_{ii(r_i-1)} & 1 & \cdots & 0\\
                \vdots & \vdots & \ddots & \vdots\\
                b_{ii2}-b_{ii1} & b_{ii3}-b_{ii2} & \cdots & 1\\
                b_{ii1} & b_{ii2} & \cdots & -1
            \end{smallmatrix}
        \end{bmatrix}, i=j,\]
    \[W_{ij}=
        \begin{bmatrix}
            \begin{smallmatrix}
                0 & 0 & \cdots & 0 & \cdots & 0\\
                -b_{ij(r_i-1)} & 0 & \cdots & 0 & \cdots & 0\\
                \vdots & \vdots & \ddots & \vdots & \ddots & \vdots\\
                b_{ij2}-b_{ij1} & b_{ij3}-b_{ij2} & \cdots & -b_{ij(r_i-1)} & \cdots & 0\\
                b_{ij1} & b_{ij2} & \cdots & b_{ij(r_i-1)} & \cdots & 0
            \end{smallmatrix}
        \end{bmatrix}, \text{if } r_i<r_j, 
    W_{ij}=\begin{bmatrix}
        \begin{smallmatrix}
            0 & 0 & \cdots & 0 & 0\\
            \vdots & \vdots & \ddots & \vdots & \vdots\\
            -b_{ij(r_j-1)} & 0 & \cdots & 0 & 0 \\
            \vdots & \vdots & \ddots & \vdots & \vdots\\
            b_{ij2}-b_{ij1} & b_{ij3}-b_{ij2} & \cdots & -b_{ij(r_j-1)} & 0\\
            b_{ij1} & b_{ij2} & \cdots & b_{ij(r_j-1)} & 0
        \end{smallmatrix}
    \end{bmatrix}, \text{if } r_i\geq r_j, \]
    \(i\neq j\), and the dimensions of \(W_{ii}\) and \(W_{ij}\) are \((r_i+1)\times r_i\) and \((r_i+1)\times r_j\), respectively. {Since the row sum of each block matrix \(W_{ij}\) (for \(i,j = 1, \cdots, p\)) is identical to \(0_{1\times r_{j}}\), the sum of all rows within the \(i\)-th subsystem block of \(Q_{1}S_{2}\) structurally collapses to \([0_{1 \times n}, \underbrace{0, \cdots, 1, \cdots, 0}_{p \text{ columns}}]\), where the unique unity element resides precisely at the \(i\)-th column index of the second vector partition.}
    

    {Given that \((Q_1 S_2)^{-1}(Q_1 S_2) = I_{n+p}\) and the sums of all rows for each subsystem of \(Q_{1}S_{2}\) match the last \(p\) rows of \(I_{n+p}\), it follows that the last \(p\) rows of \((Q_1 S_2)^{-1}\) correspond exactly to \(\text{diag}(\Gamma_1, \cdots, \Gamma_p)\), where \(\Gamma_i = [1, 1, \cdots, 1]_{1 \times (r_i+1)}\) for \(i = 1, \cdots, p\).}


    {Therefore}, the last \(p\) rows of \eqref{eq_16} are 
    \begin{equation}\label{eq_17}
        \begin{array}{r@{}l}
            V_0\left(\tilde{f}(z)-\tilde{\hat{f}}(z)\right)= & \text{diag}(\Gamma_1,\cdots,\Gamma_p)  
              \cdot \text{diag}(Z_1, \cdots, Z_p) \cdot V_0 \Delta \tilde{f}(z)\\ 
             = & \text{diag}(\Gamma_1 Z_1, \cdots, \Gamma_p Z_p)\cdot V_0 \Delta \tilde{f}(z).
        \end{array}
    \end{equation}

    {Since \(V_{0}\) is diagonal according to condition b), it commutes with \(\text{diag}(\Gamma_1Z_1,\cdots,\Gamma_pZ_p)\) in \eqref{eq_17}. Canceling the nonsingular matrix \(V_{0}\) from both sides of the equation then leads to} 
    
    \begin{equation}\label{eq:DisturbanceErrorV0diagonal}
        \tilde{f}(z)-\tilde{\hat{f}}(z)=\text{diag}(\Gamma_1 Z_1, \cdots, \Gamma_p Z_p)\cdot \Delta \tilde{f}(z).
    \end{equation}
    
    Taking the inverse \(z\)-transform of \eqref{eq:DisturbanceErrorV0diagonal} yields
    \begin{small}
    \begin{equation}\label{eq_18}
        \begin{array}{r@{}l}
            f_i(k)-\hat{f}_i(k) = & \Delta f_i(k-r_i-1)+\cdots+\Delta f_i(k-1)\\
            = & f_i(k-r_i)-f_i(k-r_i-1)+\cdots+f_i(k-1)
            -f_i(k-2)+f_i(k)-f_i(k-1)\\
            = & f_i(k)-f_i(k-r_i-1), \forall i=1,\cdots,p. 
        \end{array}
    \end{equation}
    \end{small}
    {Then}, we have \(\hat{f}_i(k)=f_i(k-r_i-1)\), \(\forall i=1,\cdots,p\) in \eqref{eq_thm1}.

    {Next,} we {show that} if Assumption \ref{assum1} does not hold, then \eqref{eq_thm1} cannot be obtained. {Assuming} \((A_0, C_0)\) is observable and \(V_0\) is nonsingular, there exists an invertible matrix \(S_1\) that {establishes the algebraic equivalence between \((A, E, C)\) and \((A_1, E_1, C_1)\), yielding a structure identical to \eqref{lem_eq1} except for the new \(\acute{E}_0=S_0E_0V_0^{-1}\).} Without loss of generality, suppose that the \(i\)-th subsystem has zero dynamics or lacks a well-defined vector relative degree. {It follows from} Assumption \ref{assum1} and Lemma \ref{lem:systemObservability} {that} the submatrix of the new \(\acute{E}_0\) corresponding to the \(i\)-th subsystem (i.e., \(\acute{E}^0_{ii}\)) {contains} non-zero entries {outside of its} last row. 
    
    {Assuming \((A_1, C_1)\) is observable to permit the derivation of \eqref{eq_thm1}.  Applying the same sequence of similarity transformations as executed in the sufficient proof,} we find that \(E^3_{ii}=((Q_1S_2)S_1E)_{ii}\) {associated with} the \(i\)-th subsystem {retains non-zero entries in postions other than its last row.} Moreover, using the same procedure to derive the last row of \((Q_1S_2)^{-1}\) corresponding to the \(i\)-th subsystem, we find that all entries in the last row have identical values. Therefore, the last row of \eqref{eq_15} associated with the \(i\)-th subsystem can be written as 
    \begin{equation}\label{eq_thm_1add}
            v^0_{ii}  \left(\tilde{f}_i(z)- \tilde{\hat{f}}_i(z)\right) = \gamma [1, 1, \cdots, 1]
             \begin{bmatrix}
                1/z & 1/z^2 & \cdots & 1/z^{r_i+1}\\
                \vdots & \vdots & \ddots & \vdots\\
                0 & 0 & \cdots & 1/z^2\\
                0 & 0 & \cdots & 1/z
            \end{bmatrix} \begin{bmatrix}
                \beta_1\\
                \beta_2\\
                \vdots \\
                \beta_{r_i+1}
            \end{bmatrix}\Delta \tilde{f}_i(z), 
    \end{equation}
    where \(v^0_{ii}\) is the entry of \(V_0\) in the \(i\)-th row and \(i\)-th column, \(\gamma\) is the identical value of the last row of \((Q_1S_2)^{-1}\) corresponding to the \(i\)-th subsystem, and \(E^3_{ii}=[\beta_1, \cdots, \beta_{r_i+1}]^{\top}\). It follows from \eqref{eq_thm_1add} that \(\hat{f}_i(k)\) is equal to a linear combination of \(f_i(k)\), \(f_i(k-1)\), \(\cdots\), \(f_i(k-r_i-1)\) rather than just a single \(f_i(k-r_i-1)\). This is a contradiction. \hfill $\square$
\end{pf}

{ 
\begin{rmk}
    For MIMO systems, there exist infinitely many observer gain matrices \(L\) capable of assigning the system eigenvalues to desired locations, as established by the dual of Lemma 2.4 in \cite[p. 336]{antsaklis2006linear}. Theorem \ref{thm2} provides a method to calculate the observer gain matrix \(L\) to simultaneously achieve complete subsystem decoupling and eigenvalue assignment. Without enforcing subsystem decoupling, the performance of the disturbance estimation degrades under time-varying disturbances because the reconstructed disturbances become contaminated by cross-channel dynamic coupling.
    
    To determine the observer gain \(L\) in \eqref{eq_3} and \eqref{eq:newMBESO}, \(\bar{L}_3\) is first calculated by decoupling all subsystems of \(A_3-\bar{L}_3\bar{C}_3\) and placing all eigenvalues at desired locations. Using \eqref{eq_8}, \eqref{eq_10}, \eqref{eq_11}, and \eqref{eq:L3_bar} yields 
    \begin{equation}\label{eq:observerGain}
        L=S_1^{-1}S_2^{-1}Q_1^{-1}\bar{L}_3 P.
    \end{equation}
\end{rmk}

\begin{rmk}
    To implement the MB-ESO within high-dimensional systems, potential numerical ill-conditioning must be carefully evaluated. 
    From \eqref{eq:observerGain}, the offline synthesis of L relies on the similarity transformations \((S_1, S_2, Q_1)\). First, \(S_1 = \text{diag}(S_0, V_0)\) is a block-diagonal matrix comprising the plant's observability matrix \(S_0\) and the disturbance gain matrix \(V_0\). Following \cite{muller1972analysis}, numerical reliability requires that the minimum eigenvalues of \(S_{0}S_{0}^{T}\) and \(V_{0}V_{0}^{T}\) remain well-bounded away from zero. Second, because \(S_2\) is the observability matrix of the augmented system, the minimum eigenvalue of \(S_{2}S_{2}^{T}\) must likewise remain sufficiently large. Finally, the nonsingular matrix \(Q_1\) transforms the system into a pseudo-observer companion form for subsystem decoupling and pole placement. Numerical case studies show that all eigenvalues of \(Q_1\) are identically equal to unity, implying that \(Q_1\) does not degrade the condition number. While observed inductively, a formal analytical proof of eigenvalues of \(Q_1\) is left for future work.
\end{rmk}

}

\begin{rmk}
    Theorem \ref{thm2} shows that the existence of an MB-ESO, like that of a conventional MIMO ESO, is fundamentally determined by the relative degree structure and the invertibility of the disturbance gain matrix. The key difference is that the MB-ESO assigns disturbances to channels with a disturbance vector relative degree and no disturbance zero dynamics, thereby fully exploiting model information and available measurements. {
    Recent studies \cite{chen2021on,chen2024quadrotor} have demonstrated that using multiple measurements together with mismatched disturbance rejection improves robustness to measurement noise, reduces noise in the control signal, and enhances tracking performance compared with conventional approaches that lump all disturbances into a single term and therefore typically require high observer gains.}
\end{rmk}

{A key structural condition in Theorem \ref{thm2} is the requirement that the disturbance gain matrix \(V_{0}\) be strictly diagonal, which is mathematically necessary to ensure that \(V_{0}\) commutes with the diagonal matrix \(\text{diag}(\Gamma_1 Z_1, \dots, \Gamma_p Z_p)\) in \eqref{eq_17} to recover accurate reconstruction of \(f(k)\). In practice, however, this diagonal condition may be violated. To circumvent this limitation, the GMB-ESO framework in \eqref{eq:newMBESO} must be deployed to reconstruct \(\hat{f}(k)\), which are rigorously established in Theorem \ref{thm:generalHighGain} below.}

\begin{theorem}\label{thm:generalHighGain}
    Condition \eqref{eq:FEstimationGeneral} holds for observer \eqref{eq:newMBESO}:
    \begin{equation}\label{eq:FEstimationGeneral}
        \hat{f}(k)=f(k-r_{\max}-1), ~\forall k>r_{\max}+1,
    \end{equation}
    if and only if a) system \eqref{eq_1} satisfies Assumption \ref{assum1}, b) all eigenvalues of the observer are placed at the origin, and c) all subsystems of the observer are decoupled. 
\end{theorem}

\begin{pf}
    The only difference between systems \eqref{eq_1} and \eqref{eq:systemRewritten} is the coefficient matrix of the disturbance vector. It is easy to verify that \(c_i^0A_0^{\kappa}E_0V_0^{-1}=0_{1\times p}\), \(\forall 1\leq i\leq p\) and \(\forall \kappa<r_i-1\), and 
    \begin{equation}\label{eq:newV0}
        V_0^{\text{new}}\triangleq\begin{bmatrix}
            c_1^0A_0^{r_1-1}E_0V_0^{-1}\\
            \vdots\\
            c_p^0A_0^{r_p-1}E_0V_0^{-1}
        \end{bmatrix}=I_{p}.
    \end{equation}
    Since \(V_0^{\text{new}}\) is diagonal, we can obtain the following result using Theorem \ref{thm2} 
    \begin{equation}\label{eq:estimateNewF}
        \hat{f}^{\text{new}}_i(k)=f_i^{\text{new}}(k-r_i-1), ~~\forall i=1,\cdots,p ~\text{and} ~k>r_i+1. 
    \end{equation}
    From \eqref{eq:delayF}, \eqref{eq:newf}, \eqref{eq:estimateNewF}, and \(f_{\text{new}}(k)=V_0 f(k)\), we have \eqref{eq:FEstimationGeneral}. \hfill $\square$
\end{pf}

\begin{rmk}
The objective of the MB-ESO is to reconstruct the unknown input \(f\) from the measured output \(y\).
In discrete-time systems, the relative degree determines the number of unit delays between the disturbance input and the measured output.
Due to causality, the disturbance reconstruction cannot be generated until the disturbance affects the output measurement.
Therefore, when \(V_0\) is diagonal, the delay of the \(i\)-th disturbance reconstruction is \(r_i+1\).
When \(V_0\) is non-diagonal, the coupling among channels implies that the delay of the disturbance reconstruction vector is determined by the maximal relative degree, i.e., \(r_{\max}+1\).
This delay reflects a fundamental limitation imposed by causality and the system structure, and therefore characterizes the best achievable estimation delay.
\end{rmk}

\begin{rmk}
For continuous-time ESO designs based on the observability canonical form, it has been shown that the state and disturbance reconstruction errors converge exponentially as the observer bandwidth \(\omega_o\) increases, both without model information except relative degree~\cite{xue2015performance} and with model information~\cite{freidovich2008performance}. 
In contrast, Theorems~\ref{thm2} and~\ref{thm:generalHighGain} establish finite-step reconstruction of the disturbance for general MIMO discrete-time linear systems by exploiting all available model information and measurements.
\end{rmk}

\subsection{Transient Behavior of the Estimation Error}
Besides finite-step convergence, it is also important to understand the transient behavior of the disturbance estimation error. 
In particular, the observer bandwidth $\omega_o$ influences the evolution of the estimation error and determines the smoothness of the reconstructed disturbance in the presence of measurement noise. 
The following theorems analyze these properties by establishing the monotonicity of the estimation error with respect to both $\omega_o$ and time $k$, and by deriving an explicit bound on the disturbance estimation error.

\begin{theorem}\label{thm3}
    If Assumption \ref{assum1} holds for system \eqref{eq_1} and \(V_0\) is diagonal, then the MB-ESO \eqref{eq_3} has the following properties.  

    (i) The estimation error of \(f\) is 
    \begin{equation}
        f(k)-\hat{f}(k)=\text{diag}(h_1(k),\cdots,h_p(k))\ast \Delta f(k), 
    \end{equation}
    where \(\ast\) represents convolution, \(\Delta f(k) = f(k+1) - f(k)\) is the variation of the true disturbance, 
    and\footnote{For simplicity, let \(\prod_{j=0}^{-1}(k+j)=1\).}
    \begin{equation}\label{eq_27}
        \begin{array}{l}
            h_i(k)=  
            \begin{cases}
                1 & 1\leq k \leq r_i+1,\\
                    \sum\limits_{\iota = 1}^{r_i+1}  \dfrac{1}{(\iota-1)!}\bigg(\prod\limits_{j=-\iota+1}^{-1}(k
                     +j)\bigg)(1-\omega_{o_i})^{\iota-1}\omega_{o_i}^{k-\iota} 
                & k\geq r_i+2.
            \end{cases} 
        \end{array}
    \end{equation}


    (ii) After \(r_i+1\) steps following a change in \(f_i\), the estimation error of \(f_i\) decreases monotonically to zero as \(\omega_{o_i}\) decreases, where \(0<\omega_{o_i}<1\).

    (iii) For any fixed \(0<\omega_{o_i}<1\), the estimation error of \(f_i\) decreases monotonically to zero as time \(k\) increases, after \(r_i+1\) steps following a change in \(f_i\).
\end{theorem}

\begin{pf}
    (i) After a series of similarity transformations made in Theorem \ref{thm2}, the estimation error dynamics of observer \eqref{eq_3} for the augmented system \eqref{eq_2} can be written as
    \begin{equation}\label{eq_19}
        e_3(k+1)=(A_3-L_3C_3)e_3(k)+E_3 \Delta f(k),
    \end{equation}
    where \(e_3(k)=(Q_1S_2S_1)e(k)\), \(L_3=(Q_1S_2S_1)L\), \(A_3=(Q_1S_2S_1)A(Q_1S_2S_1)^{-1}\), \(C_3=C(Q_1S_2S_1)^{-1}\), and \(E_3=(Q_1S_2S_1)E\). {According to \eqref{eq:L3_bar} and \eqref{eq:L3_bar_zero}, the observer gain matrix \(L_{3}\) can be computed such that all off-diagonal submatrices of \(A_3-L_3C_3\) are zero. Concurrently, the eigenvalues of the diagonal submatrices, \((A_3-L_3C_3)_{ii}\) for \(i=1,\cdots,p\), are assigned to \(\omega _{o_{i}}\), respectively. The entries in the first column of \((A_3-L_3 C_3)_{ii}\) are then determined by the following matrix:}
    \begin{equation}\label{eq_20}
        (A_3-L_3C_3)_{ii}=\begin{bmatrix}
            (-1)^2\binom{r_i+1}{1}\omega_{o_i} & 1 & \cdots & 0 & 0\\
            (-1)^3\binom{r_i+1}{2}\omega_{o_i}^2 & 0 & \cdots & 0 & 0\\
            \vdots & \vdots & \ddots & \vdots & \vdots\\
            (-1)^{r_i+1}\binom{r_i+1}{r_i}\omega_{o_i}^{r_i} & 0 & \cdots & 0 & 1\\
            (-1)^{r_i+2}\binom{r_i+1}{r_i+1}\omega_{o_i}^{r_i+1} & 0 & \cdots & 0 & 0\\
        \end{bmatrix}. 
    \end{equation}
    Now we need to obtain the transfer function of \eqref{eq_19} by taking \(\Delta f(k)\) as input and the last \(p\) rows of \(e(k)\), \emph{i.e.}, \(f(k)-\hat{f}(k)\), as output. Moreover, only the subsystems on the diagonal need to be considered because all subsystems are decoupled by the observer gain \(L_3\). According to \eqref{eq_20}, it is not easy to calculate \((zI-(A_3-L_3C_3)_{ii})^{-1}\) directly. A similarity transformation \(Q_2\) is used to transform \(A_3-L_3 C_3\) to a controller companion form \(A_4\), where \(A_4=[A^4_{ij}]\), \(i,j=1,\cdots,p\), \(A^4_{ii}=\)
    \(
        \begin{bmatrix}\begin{smallmatrix}
            0 & 1 & \cdots & 0\\
            \vdots & \vdots & \ddots & \vdots\\
            0 & 0 & \cdots & 1\\
            (-1)^{r_i+2}\binom{r_i+1}{r_i+1}\omega_{o_i}^{r_i+1} & (-1)^{r_i+1}\binom{r_i+1}{r_i}\omega_{o_i}^{r_i} & \cdots & (-1)^2\binom{r_i+1}{1}\omega_{o_i}
            \end{smallmatrix}
        \end{bmatrix}, i=j,
    \)
    \(A^4_{ij}=0_{(r_i+1)\times (r_j+1)}\), \(i\neq j\), \(Q_2=[Q^2_{ij}]\), \(i,j=1,\cdots,p\), 
    \(
        Q^2_{ii}=\begin{bmatrix}\begin{smallmatrix}
            1 & 0 & \cdots & 0\\
            (-1)\binom{r_i+1}{1}\omega_{o_i} & 1 & \cdots & 0\\
            \vdots & \vdots & \ddots & \vdots\\
            (-1)^{r_i-1}\binom{r_i+1}{r_i-1}\omega_{o_i}^{r_i-1} & (-1)^{r_i-2}\binom{r_i+1}{r_i-2}\omega_{o_i}^{r_i-2} & \cdots & 0\\
            (-1)^{r_i}\binom{r_i+1}{r_i}\omega_{o_i}^{r_i} & (-1)^{r_i-1}\binom{r_i+1}{r_i-1}\omega_{o_i}^{r_i-1} & \cdots & 1 
            \end{smallmatrix}
        \end{bmatrix} 
    \), \(i=j\), \(Q^2_{ij}=0_{(r_i+1)\times (r_j+1)}\), \(i\neq j\). 
    
    Then, \eqref{eq_19} can be reformulated as
    \begin{equation}\label{eq_21}
        e_4(k+1)=A_4e_4(k)+E_4 \Delta f(k),
    \end{equation}
    where \(e_4(k)=Q_2^{-1}e_3(k)\), \(A_4=Q_2^{-1}(A_3-L_3C_3)Q_2\), and 
    \(E_4=Q_2^{-1}E_3\).

    Taking the \(z\)-transform of \eqref{eq_21} without considering the initial condition yields
    \begin{equation}\label{eq_22}
        \tilde{e}_4(z)=(zI-A_4)^{-1}E_4\Delta \tilde{f}(z).
    \end{equation}
    Since \(A_4\) is a diagonal matrix in a controller companion form, \(E_4=[\underline{e}^4_{ij}]\), \(i,j=1,\cdots,p\), \(\underline{e}^4_{ii}=[0, \cdots, 0, v^0_{ii}]_{1\times (r_i+1)}^{\top}\), \(i=j\), and \(\underline{e}^4_{ij}=0_{(r_i+1)\times 1}\), \(i\neq j\), the transfer function of \eqref{eq_22} is also a diagonal matrix, where the \(i\)-th submatrix on the diagonal is
    \begin{equation}\label{eq_23}
        ((zI-A_4)^{-1}E_4)_{ii}=\Lambda_i v^0_{ii}, 
    \end{equation}
    and \(\Lambda_i=[1/(z-\omega_{o_i})^{r_i+1}, z/(z-\omega_{o_i})^{r_i+1}, \cdots, z^{r_i}/(z-\omega_{o_i})^{r_i+1}]^{\top}\). From \eqref{eq_19}, \eqref{eq_21}, \eqref{eq_22}, and \eqref{eq_23}, 
    the \(z\)-transform of the original observer error dynamics is 
    \begin{equation}\label{eq_24}
        S_1\tilde{e}(z)=(Q_1S_2)^{-1}Q_2(zI-A_4)^{-1}E_4\Delta \tilde{f}(z). 
    \end{equation}
    {According to \eqref{eq_17} and \eqref{eq_23}}, the last \(p\) rows of \eqref{eq_24} are 
    \begin{equation}\label{eq_25}
            V_0 (\tilde{f}(z)-\tilde{\hat{f}}(z))
            = \text{diag}(\Gamma_1,\cdots,\Gamma_p) Q_2 \text{diag}(\Lambda_1, \cdots, \Lambda_p) V_0\Delta \tilde{f}(z). 
    \end{equation}
    Then, the transfer function of \eqref{eq_25} for the \(i\)-th subsystem is
    \begin{equation}\label{eq_26}
        \begin{array}{r@{}l}
            \tilde{h}_i(z) & =[1, \cdots, 1, 1]Q^2_{ii}\begin{bmatrix}
                1/(z-\omega_{o_i})^{r_i+1}\\
                \vdots\\
                z^{r_i-1}/(z-\omega_{o_i})^{r_i+1}\\
                z^{r_i}/(z-\omega_{o_i})^{r_i+1}
            \end{bmatrix}\\
            & =\sum\limits_{\iota = 1}^{r_i+1} \dfrac{(1-\omega_{o_i})^{\iota-1}(z-\omega_{o_i})^{r_i+1-\iota}}{(z-\omega_{o_i})^{r_i+1}}. 
        \end{array}
    \end{equation}
    Taking the inverse \(z\)-transform of \eqref{eq_26} 
    yields \eqref{eq_27}. 

    (ii) {Because the original system is linear, it satisfies additivity and homogeneity. Consequently, if the transfer functions in \eqref{eq_27} decrease as \(\omega _{o}\) decreases for all \(k\geq 0\), the estimation error of the disturbance \(f\) will likewise decrease with \(\omega _{o}\) for all \(k\geq 0\).}

    The derivative of \(h_i(k)\) with respect to \(\omega_{o_i}\) is 
    \begin{equation}\label{eq_28}
        \dfrac{\partial h_i(k)}{\partial \omega_{o_i}} = \dfrac{1}{r_i!}\left(\prod\limits_{j=-r_i-1}^{-1}(k+j)\right)(1-\omega_{o_i})^{r_i}\omega_{o_i}^{k-r_i-2}.
    \end{equation}
    Therefore, if \(0<\omega_{o_i} <1\), then \(\partial h_i/\partial \omega_{o_i} >0\) for all \(k>r_i+1\). {Given that \(h_i(k)>0\) for all \(k>r_i+1\), the sequence \(h_i(k)\) decreases monotonically to zero with respect to \(\omega _{o_{i}}\) after \(r_i+1\) steps following a change in \(f_i\).}

    (iii) From \eqref{eq_27}, we have
    \begin{equation}\label{eq:timeDecreaseOriginal}
            h_i(k+1)-h_i(k)=  -\dfrac{1}{r_i !}\left(\prod\limits_{j=-r_i}^{-1}(k+j)\right) 
              \cdot (1-\omega_{o_i})^{r_i+1}\omega_{o_i}^{k-r_i-1}
    \end{equation}
    Therefore, if \(0<\omega_{o_i} <1\), then \(h_i(k+1)-h_i(k)<0\) for all \(k\geq r_i+1\). {Since \(h_i(k)>0\) for all \(k>r_i+1\),} \(h_i(k)\) decreases monotonically {to zero} with respect to time \(k\) {after \(r_i+1\) steps following a change in \(f_i\)}. \hfill $\square$
\end{pf}

{
\begin{rmk}

Recent studies \cite{zhang2024disturbance,zhang2026dual} demonstrate the potential to learn a prior disturbance model from ESO-based disturbance reconstruction and incorporate it into the system model to improve performance. As shown in \cite{zhang2024disturbance}, a learning-enabled ESO can ultimately outperform a baseline ESO based on an identified system model. Reference \cite{zhang2026dual} further investigates a critical question: although the \emph{benefits} of incorporating model information are well recognized, how can the associated \emph{risk} be controlled when the model is inaccurate? In \cite{zhang2026dual}, a Lipschitz neural network is employed to balance model accuracy and robustness. We would also like to emphasize that the theoretical foundation established in this work is critical to such data-driven approaches. Well-characterized properties, monotonic convergence and bounded reconstruction error, ensure that disturbance learning (essentially supervised learning using ESO-reconstructed disturbances as labels) is built on a solid foundation.
\end{rmk}
}

The following result shows the properties of the GMB-ESO \eqref{eq:newMBESO} when \(V_0\) is not diagonal. 

\begin{theorem}\label{thm:generalErrorDynamic}
    If Assumption \ref{assum1} holds for system \eqref{eq_1}, then the estimation error of \(f\) of the GMB-ESO \eqref{eq:newMBESO} is
    \begin{equation}\label{eq:newTF_f}
        f(k)-\hat{f}(k)=V_0^{-1}\text{diag}(h^{\text{new}}_1(k),\cdots,h^{\text{new}}_p(k))\ast\Delta f_{\text{new}}(k), 
    \end{equation}
    where \(\Delta f_{\text{new}}(k) = f_{\text{new}}(k+1) - f_{\text{new}}(k)\), and 
    \begin{equation}\label{eq:newh_i}
            h^{\text{new}}_i(k)=  
            \begin{cases}
                1 & 1\leq k \leq r_{\text{max}}+1,\\
                    \sum\limits_{\iota = 1}^{r_i+1}  \dfrac{1}{(\iota-1)!}\bigg(\prod\limits_{j=-\iota+1}^{-1}(k-(r_{\text{max}}-r_i)
                    +j)\bigg)(1-\omega_{o_i})^{\iota-1}\omega_{o_i}^{k-(r_{\text{max}}-r_i)-\iota} 
                & k\geq r_{\text{max}}+2.
            \end{cases} 
    \end{equation}
\end{theorem}

\begin{pf}
    After a series of similarity transformations made in Theorem \ref{thm3}, the estimation error dynamics of the new observer \eqref{eq:newMBESO} for the newly augmented system \eqref{eq:newAugmentedSystem} can be written like \eqref{eq_22} in the \(z\)-domain
    \begin{equation}
        \tilde{e}_4(z)=(zI-A_4)^{-1}E_4\Delta \tilde{f}_{\text{new}}(z),
    \end{equation}
    where \(A_4\) {remains} the same diagonal matrix, but {the entries of \(E_4\) satisfy \(v^0_{ii}=1\) because} \(V_0^{\text{new}}=I_p\) in \eqref{eq:newV0}. 

    The last \(p\) rows of the observer error dynamics of \eqref{eq:newMBESO} are 
    \begin{equation}\label{eq:newEstimatioError}
            \tilde{f}_{\text{new}} (z)-\tilde{\hat{f}}_{\text{new}}(z)
            =  \text{diag}(\Gamma_1,\cdots,\Gamma_p) Q_2 \text{diag}(\Lambda_1, \cdots, \Lambda_p) \Delta \tilde{f}_{\text{new}}(z). 
    \end{equation}
    The inverse \(z\)-transforms of \eqref{eq:newEstimatioError} are 
    \begin{equation}
        f_i^{\text{new}}(k)-\hat{f}_i^{\text{new}}(k)=h_i(k)\ast\Delta f_i^{\text{new}}(k), ~i=1,\cdots,p, 
    \end{equation}
    where \(h_i(k)\) is \eqref{eq_27}. {Synchronizing} \(\hat{f}_{\text{new}}(k)\) as in \eqref{eq:delayF} yields
    \begin{equation}\label{eq:TF_fnew}
        f_i^{\text{new}}(k)-\hat{\bar{f}}^{\text{new}}_i(k)=h_i^{\text{new}}(k)\ast\Delta f_i^{\text{new}}(k), ~i=1,\cdots,p, 
    \end{equation}
    where \(h^{\text{new}}_i(k)\) is given by \eqref{eq:newh_i}. 
    
    It follows from \eqref{eq:systemRewritten}, \eqref{eq:newf}, and \eqref{eq:TF_fnew} that the estimation error of \(f\) is given by \eqref{eq:newTF_f}. \hfill $\square$
\end{pf}

\begin{rmk}
Theorem \ref{thm3} shows that the estimation error of the disturbance \(f\) decreases monotonically with respect to \(\omega_o\) and \(k\). 
Consequently, the estimation error during and after a change in \(f\) cannot increase when \(\omega_o\) decreases or when \(k\) increases. 
This property implies that the reconstructed disturbance exhibits no overshoot (as observed in the simulations of \cite{bakhshande2015proportional,li2012generalized}) and ensures reliable estimation performance of the MB-ESO \eqref{eq_3} in both transient and steady states. 
However, Theorem \ref{thm:generalErrorDynamic} demonstrates that the reconstructed \(\hat{f}\) produced by the GMB-ESO \eqref{eq:newMBESO} does not possess this monotonicity property due to the presence of \(V_0^{-1}\) in \eqref{eq:newTF_f}. {Nonetheless, as established in Theorem \ref{thm:generalHighGain}, the disturbance can still be reconstructed when the observer bandwidth approaches infinity.}
\end{rmk}

\begin{rmk}
    Theorem~\ref{thm3} establishes an explicit bound on the disturbance estimation error, whereas \cite{freidovich2008performance,xue2015performance} provide only conservative bounds. This bound is critical for robust optimal control in safety-critical systems \cite{alan2022disturbance,chen2023robust}. Notebly, the bound depends solely on $\omega_o$, $\Delta f$, and the disturbance relative degree.
\end{rmk}

\begin{rmk}
    The transfer function $\tilde{h}_i(z)$ in~\eqref{eq_26} can be interpretated as a combination of low-pass filters, which explains the smoothing effect of the MB-ESO in the presence of measurement noise.
\end{rmk}

\section{Numerical Simulations and Discussion}\label{sec:simulation}
This section demonstrates the properties and effectiveness of the proposed MB-ESO and GMB-ESO through six simulation studies. Simulation \ref{sim1} verifies the finite-step convergence property in Theorem~\ref{thm:generalHighGain}. Simulation \ref{sim2} verifies the transient behavior of the estimation error predicted by the theoretical analysis. Simulation \ref{sim3} compares the proposed observers with existing disturbance observers. Simulation \ref{sim4} investigates the impact of invariant zeros on disturbance estimation performance. Simulation \ref{sim5} compares the reconstructed disturbances using different discretization methods, with and without invariant zeros. Simulation \ref{sim6} explains the mechanism behind the high sensitivity to high-frequency measurement noise by using a D-UIO, whose observer error dynamics contains no uncertainty term, as a benchmark.

\subsection{Verification of the Finite-Step Convergence}\label{sim1}
Consider a discrete-time linear time-invariant system of the form \eqref{eq_1} with the following coefficient matrices
\begin{equation*}
    A_0=\begin{bmatrix}
        \begin{smallmatrix}
            -0.24 & 1.88 & -0.28 & -0.016 & 0 & 0 & -0.08 & 0.04 & -0.04\\
        -0.12 & 0.44 & -0.14 & -0.008 & 0 & 0 & -0.04 & 0.02 & -0.02\\
        0 & 0 & 0 & 1 & 0.8 & 0 & 0 & 0 & 0\\
        0 & 0 & 0 & 0 & 1 & 0 & 0 & 0 & 0\\
        0 & 0 & 0 & 0 & 0 & 1 & 0 & 0 & 0\\
        -0.5 & 0.8 & -0.1 & -0.1 & -0.6 & -0.4 & -0.8 & -0.14 & 0.04\\
        0 & 0 & 0 & 0 & 0 & 0 & 0 & 1 & -1\\
        -0.4 & 0.6 & -0.1 & -0.07 & -0.4 & 0 & -0.2 & -0.1 & 0.4\\
        -0.4 & 0.6 & -0.1 & -0.07 & -0.4 & 0 & -0.2 & -0.1 & -0.6
        \end{smallmatrix}
    \end{bmatrix},
\end{equation*}
\begin{equation*}
    E_0=\begin{bmatrix}
        4 & 2 & 0 & 0 & 0 & 0.35 & 0 & 0.4 & 0.4\\
        1 & 0.5 & 0 & 0 & 0 & 1.5 & 0 & 0.6 & 0.6\\
        0.5 & 0.25 & 0 & 0 & 0 & 0.5 & 0 & 0.5 & 0.5
    \end{bmatrix}^{\top},
\end{equation*}
\begin{equation*}
    C_0=\begin{bmatrix}
        1 & -2 & 0 & 0 & 0 & 0 & 0 & 0 & 0\\
        0 & 0 & 1 & -0.8 & 0 & 0 & 0 & 0 & 0 \\
        0 & 0 & 0 & 0 & 0 & 0 & 1 & 0 & 0
    \end{bmatrix},
\end{equation*}
and \(B_0\) is omitted due to the neglect of \(u\), without loss of generality. This system has a vector relative degree \(\{2,4,3\}\) between the disturbances and the outputs, and 
\begin{equation*}
    V_0=\begin{bmatrix}
        2 & 0.5 & 0.25\\
        0.35 & 1.5 & 0.5\\
        0.4 & 0.6 & 0.5
    \end{bmatrix}
\end{equation*}
is not diagonal. Thus, the GMB-ESO in \eqref{eq:newMBESO} is used to reconstruct disturbances. {According to \eqref{eq:observerGain}, the observer gain \(L\) is designed to fully decouple all subsystems while placing all eigenvalues at the origin. }

\begin{figure}[htbp]
    \begin{center}
        \includegraphics[width=3.5 in]{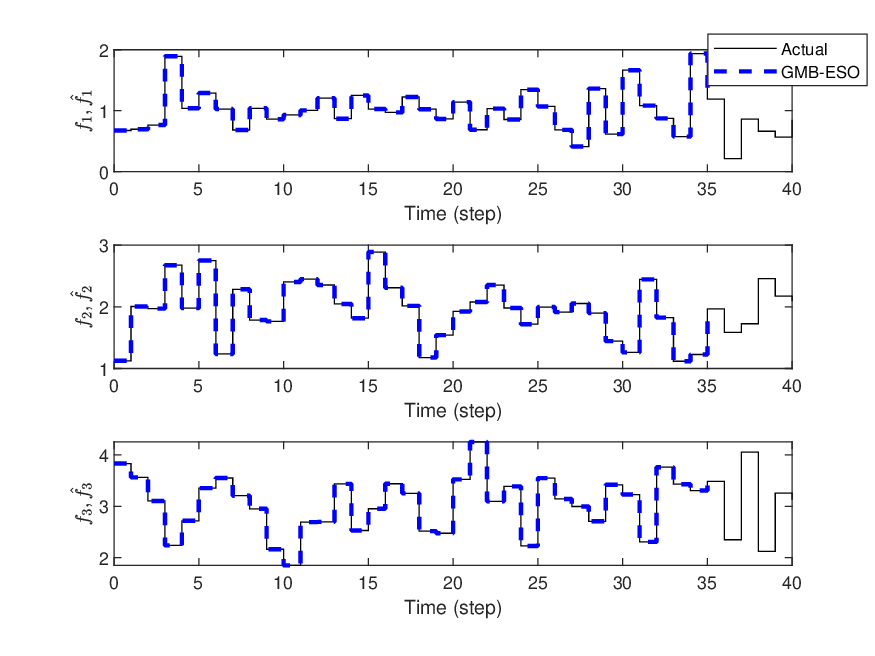}
        \caption{Disturbance estimation using the GMB-ESO. The reconstructions coincide with the true disturbances with a delay of 5 steps (Theorem~\ref{thm:generalHighGain}). The reconstructions are shifted forward by 5 steps for visual comparison.}
        \label{fig:MB_ESO_highGain}
    \end{center}
\end{figure}

Three unknown disturbances that act on the system are generated from three Gaussian distributions with means \(1\), \(2\), \(3\) and variances \(0.1\), \(0.2\), \(0.3\), respectively. The initial state of the system is \(x_0=[1, 3, 1, 5, 7, 2, 0, 1, 4]^{\top}\).  As shown in Fig. \ref{fig:MB_ESO_highGain}, the reconstructed disturbances using the GMB-ESO are all equal to the true disturbances with delay \(5\), which confirms Theorem \ref{thm:generalHighGain}. The reconstructed disturbances have been shifted forward by 5 steps to eliminate the effect of the delays for clear visual comparison. 

\subsection{Transient Behavior of the Estimation Error}\label{sim2}
The same system as in Simulation \ref{sim1} is used to illustrate the monotonic behavior of the disturbance estimation error under the MB-ESO. Since \(V_0\) is non-diagonal, the GMB-ESO in \eqref{eq:newMBESO} is {deployed to reconstruct disturbances. As detailed in \eqref{eq:systemRewritten}, an algebraic transformation is introduced to construct a virtual disturbance vector \(f_{\text{new}}\) that maps to a decoupled, diagonalized disturbance gain matrix \(V_{0}^{\text{new}}\). Under this transformed representation, the reconstructed virtual disturbances \(\hat{f}_{\text{new}}\) retain the strict error monotonicity property proven in Theorem \ref{thm3}. Subsequently, this virtual estimation result is mapped back to the physical domain via \eqref{eq:delayF} and \eqref{eq:newf} to reconstruct the original disturbance vector \(f\). Crucially, due to this reverse mapping, the physical tracking error loses its structural monotonicity, as formalized in Theorem \ref{thm:generalErrorDynamic}.} The observer gain \(L\) is computed via \eqref{eq:observerGain} to decouple all subsystems and place all eigenvalues at \(\omega_o\). 

\begin{figure}[htbp]
    \begin{center}
        \includegraphics[width=3.5 in]{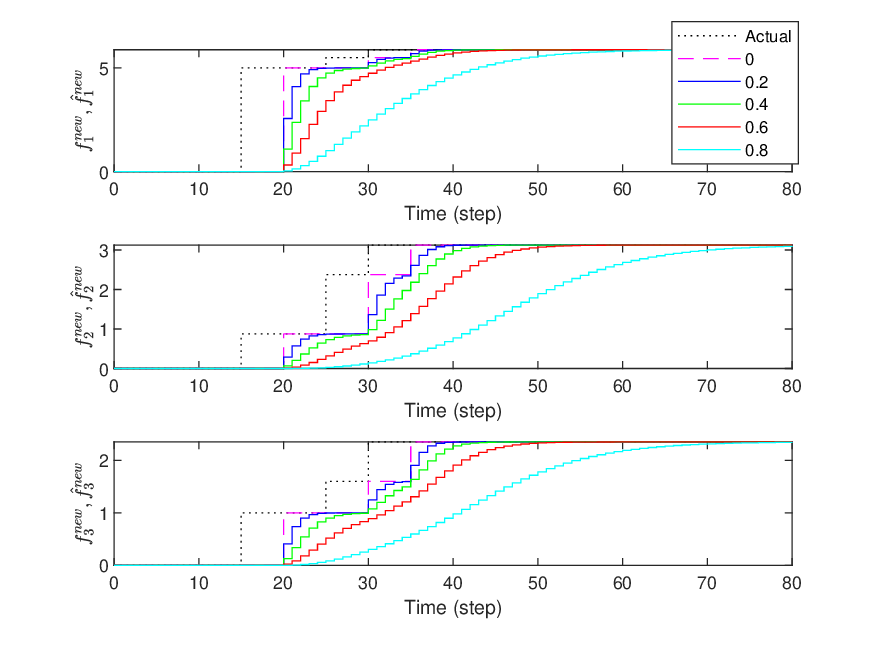}
        \caption{The reconstructed disturbances \(\hat{f}_{\text{new}}\) using the GMB-ESO in \eqref{eq:newMBESO} with different eigenvalues \(\omega_o=0\), \(\omega_o=0.2\), \(\omega_o=0.4\), \(\omega_o=0.6\), and \(\omega_o=0.8\).}
        \label{fig:ESO_new_monoto}
    \end{center}
\end{figure}

\begin{figure}[htbp]
    \begin{center}
        \includegraphics[width=3.5 in]{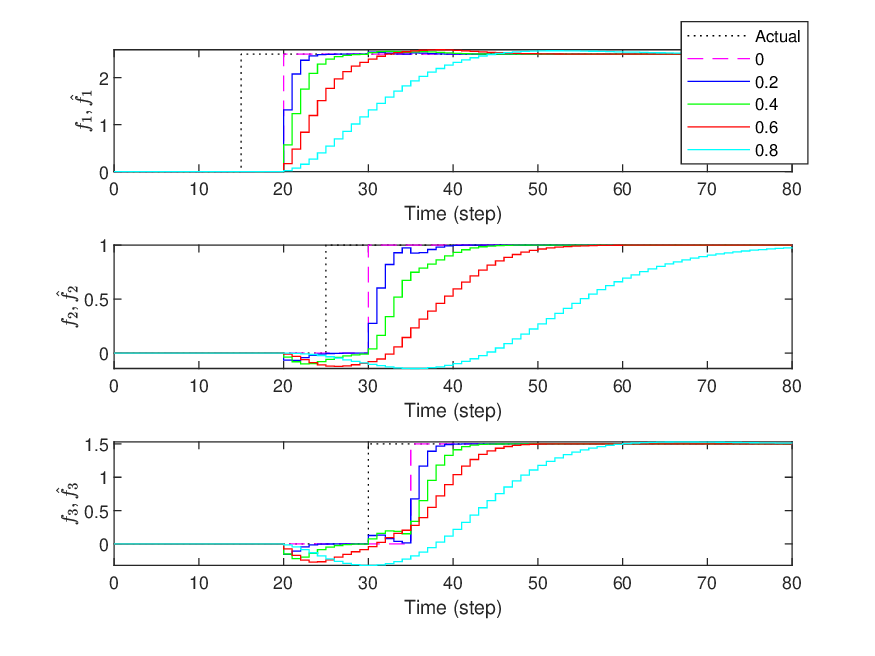}
        \caption{The reconstructed original disturbances \(\hat{f}\) converted from the GMB-ESO using \eqref{eq:newf} with different eigenvalues \(\omega_o=0\), \(\omega_o=0.2\), \(\omega_o=0.4\), \(\omega_o=0.6\), and \(\omega_o=0.8\).}
        \label{fig:ESO_monoto}
    \end{center}
\end{figure}

Step disturbance signals \(f_1\), \(f_2\), and \(f_3\) are applied to the system at the steps \(15\), \(25\), and \(30\), with corresponding magnitudes of \(2.5\), \(1\), and \(1.5\), respectively. The initial state of the system is \(x_0=[0, 0, 0, 0, 0, 0, 0, 0, 0]^{\top}\). As shown in Fig. \ref{fig:ESO_new_monoto}, the estimation error of \(f_i^{\text{new}}\), \(i=1,2,3\) decreases monotonically to zero after \(5\) steps of the change of \(f_i^{\text{new}}\) with respect to \(\omega_o\) and time \(k\), which confirms the monotonicity in Theorem \ref{thm3}. {Note that the monotonicity guarantees the absence of overshoot in the disturbance estimation.} However, the estimation error of the original disturbances \(f_i\), \(i=1,2,3\) in Theorem \ref{thm:generalErrorDynamic} does not have this monotonicity property in Fig. \ref{fig:ESO_monoto}. 

\subsection{Comparison with Existing Disturbance Observers}\label{sim3}
Conventional ESO and EHGO typically handle a MIMO system by decomposing it into multiple SISO subsystems, each possessing a cascade integrator structure. In contrast, the proposed MB-ESO and the PIO are designed directly for MIMO systems. In the following, these observers are compared with the EHGO constructed for multiple SISO subsystems \cite{wang2015output} using the following system in normal form:
\begin{equation*}
    A_0=\begin{bmatrix}
        \begin{smallmatrix}
        0 & 1 & 0 & 0 & 0 & 0 & 0 & 0 & 0\\
        -0.12 & 0.2 & -0.14 & -0.12 & 0 & 0 & -0.04 & 0.02 & 0\\
        0 & 0 & 0 & 1 & 0 & 0 & 0 & 0 & 0\\
        0 & 0 & 0 & 0 & 1 & 0 & 0 & 0 & 0\\
        0 & 0 & 0 & 0 & 0 & 1 & 0 & 0 & 0\\
        -0.5 & -0.2 & -0.1 & -0.18 & -0.6 & -0.4 & -0.8 & -0.14 & -0.1\\
        0 & 0 & 0 & 0 & 0 & 0 & 0 & 1 & 0\\
        0 & 0 & 0 & 0 & 0 & 0 & 0 & 0 & 1\\
        -0.4 & -0.2 & -0.1 & -0.15 & -0.4 & 0 & -0.2 & -0.1 & -0.7
    \end{smallmatrix}
    \end{bmatrix},
\end{equation*}
\begin{equation*}
    E_0=\begin{bmatrix}
        0 & 2 & 0 & 0 & 0 & 0 & 0 & 0 & 0\\
        0 & 0 & 0 & 0 & 0 & 1.5 & 0 & 0 & 0\\
        0 & 0 & 0 & 0 & 0 & 0 & 0 & 0 & 2
    \end{bmatrix}^{\top}, 
    C_0=\begin{bmatrix}
        1 & 0 & 0 & 0 & 0 & 0 & 0 & 0 & 0\\
        0 & 0 & 1 & 0 & 0 & 0 & 0 & 0 & 0\\
        0 & 0 & 0 & 0 & 0 & 0 & 1 & 0 & 0
    \end{bmatrix}. 
\end{equation*}
The disturbance relative vector degree of this system is \(\{2,4,3\}\), and 
\(V_0=\text{diag}(2, 1.5, 2)\)
is diagonal. Thus, the MB-ESO in \eqref{eq_3} is used to reconstruct disturbances. Step disturbance signals \(f_1\), \(f_2\), and \(f_3\) are applied to the system at the steps \(15\), \(25\), and \(30\), with corresponding magnitudes of \(1\), \(2.5\), and \(1.5\), respectively. The initial state of the system is \(x_0=[0, 0, 0, 0, 0, 0, 0, 0, 0]^{\top}\). All eigenvalues of the MB-ESO and the EHGO are placed at \(0.2\). The observer gain \(L\) of the PIO is designed using LQR method with \(Q=\text{diag}(1,1,1,1,1,1,1,1,1,\bar{q},\bar{q},\bar{q}),R=\text{diag}(\bar{r},\bar{r},\bar{r}), \bar{q}=10^4, \bar{r}=4\times 10^4\) \cite{bakhshande2015proportional}. {It should be noted that the PIO relies on the LQR method rather than eigenvalue assignment. While conventional pole placement cannot ensure subsystem decoupling, the LQR framework addresses this limitation by optimization. Ensuring this decoupling is essential, as its absence leads to severe transient oscillations and degraded tracking performance.} 

\begin{figure}[htbp]
    \begin{center}
        \includegraphics[width=3.5 in]{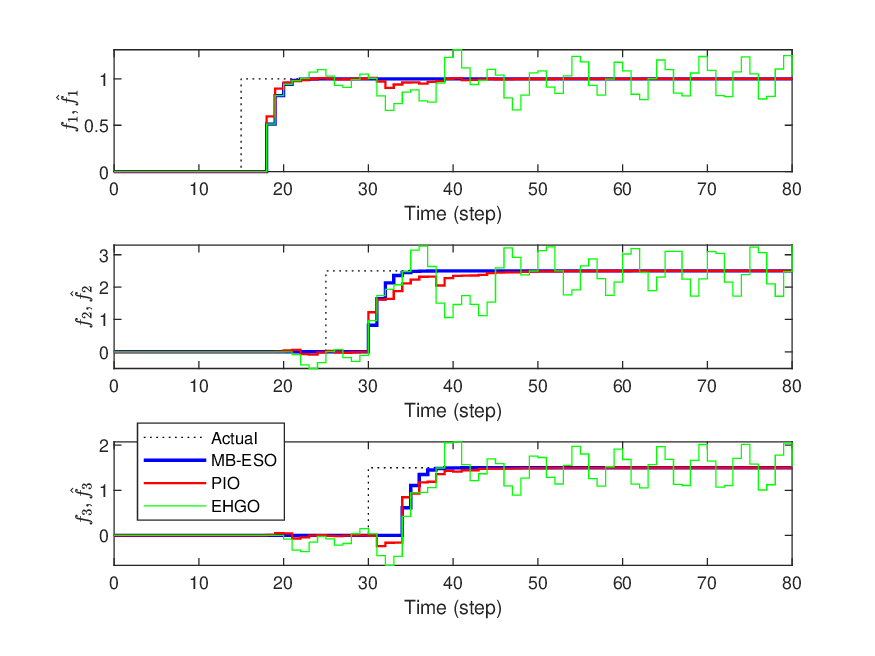}
        \caption{The reconstructed disturbances using the MB-ESO, the PIO, and the EHGO with all model information and measurements.}
        \label{fig:PIO_MBESO_EHGO}
    \end{center}
\end{figure}

As shown in Fig.~\ref{fig:PIO_MBESO_EHGO}, the EHGO performs the worst, with reconstructions fluctuating around the true disturbances because it effectively reconstructs the sum of the disturbance and the state estimation error. In contrast, the MB-ESO and the PIO achieve higher accuracy since the disturbance is the only quantity being reconstructed. The transient disturbance reconstructions of the MB-ESO and the PIO are similar; however, the PIO exhibits noticeable fluctuations because the subsystems of the MB-ESO are fully decoupled, whereas the LQR-based design cannot completely decouple them. These fluctuations become even more significant when \(V_0\) is non-diagonal.

\subsection{Influence of Invariant Zeros on Disturbance Estimation}\label{sim4}
Many disturbance observers developed within the state space framework, such as \cite{johnson1971accomodation,soffker1995state,li2012generalized}, do not require Assumption \ref{assum1}. In this example, we demonstrate how invariant zeros affect the disturbance estimation performance of the {PIO}. {Notably, when Assumption \ref{assum1} is omitted, our MB-ESO becomes structurally equivalent to both the PIO and the GESO. Consequently, the simulation results presented in this subsection are directly applicable to the GESO and MB-ESO frameworks as well.} Specifically, consider a SISO two-mass spring system characterized by the following state-space matrices:
\begin{equation*}
    A_0=\begin{bmatrix}
        0 & 1 & 0 & 0\\
        b_{110} & b_{111} & b_{120} & 0\\
        0 & 0 & 0 & 1\\
        b_{210} & 0 & b_{220} & b_{221}
    \end{bmatrix}, 
    E_0=\begin{bmatrix}
        0\\
        1\\
        0\\
        0
    \end{bmatrix}, 
    C_0=\begin{bmatrix}
        1 & 0 & 0 & 0
    \end{bmatrix}. 
\end{equation*}
The numerator of the transfer function of this system is \(z^2 - b_{221}z - b_{220}\). Thus, the zeros of the system can be placed by adjusting the coefficients \(b_{220}\) and \(b_{221}\). In the simulation, the coefficients are chosen as \(b_{110}=-0.264\), \(b_{111}=1.03\), \(b_{120}=0.02\), and \(b_{210}=0.02\). 

\begin{figure}[htbp]
    \begin{center}
        \includegraphics[width=3.5 in]{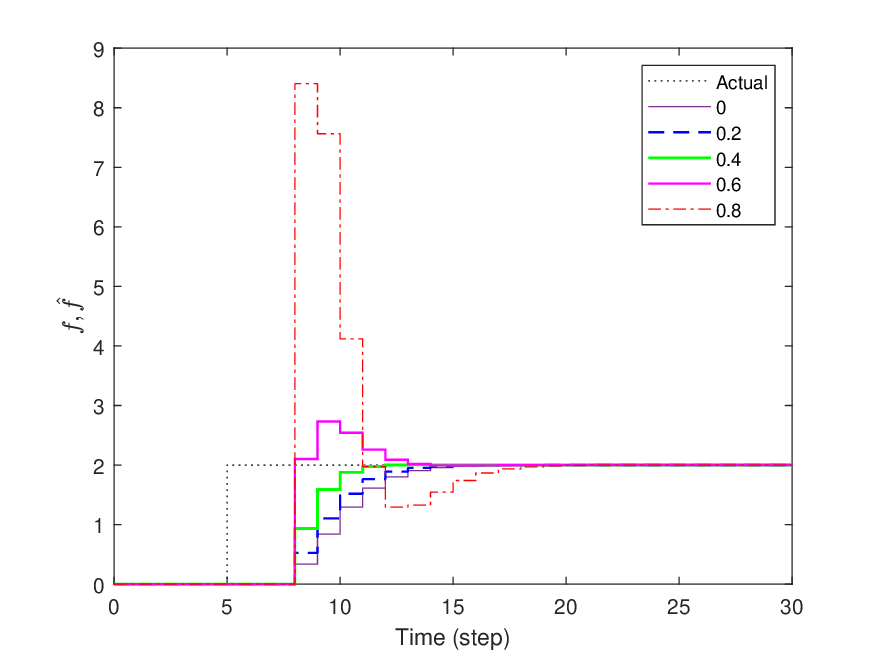}
        \caption{The reconstructed disturbances using the {PIO} with all observer eigenvalues at \(0.3\). Both zeros of the system are placed at \(0\), \(0.2\), \(0.4\), \(0.6\), and \(0.8\), respectively.}
        \label{fig:Different_Zeros}
    \end{center}
\end{figure}

\begin{figure}[htbp]
    \begin{center}
        \includegraphics[width=3.5 in]{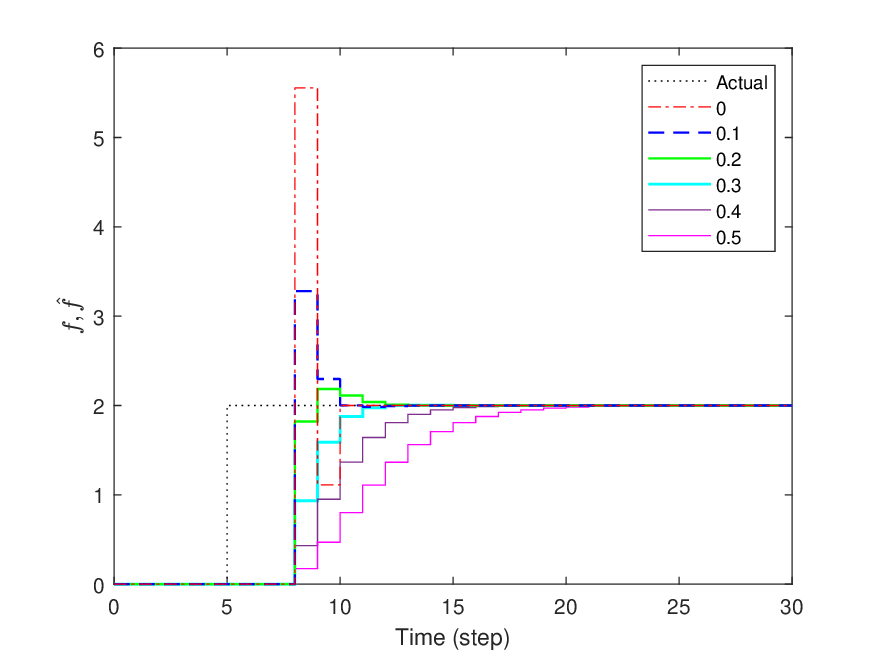}
        \caption{The reconstructed disturbances using the {PIO} with all observer eigenvalues at \(0\), \(0.1\), \(0.2\), \(0.3\), \(0.4\), and \(0.5\), respectively. Both zeros of the system are placed at \(0.4\).}
        \label{fig:Different_wo_zero}
    \end{center}
\end{figure}

A step disturbance $f$ of magnitude $2$ is applied at step $5$, with initial state $x_0=[0,0,0,0]^{\top}$. As shown in Fig.~\ref{fig:Different_Zeros}, reconstructed disturbance fluctuations increase as zeros approach $1$, while no fluctuation occurs if both zeros are below $0.4$. This is because the  {PIO with low observer bandwidth} behaves like a low-pass filter, smoothing overshoot caused by small zeros. However, Fig.~\ref{fig:Different_wo_zero} shows that placing the observer eigenvalues closer to the origin improves estimation but leads to fluctuations caused by invariant zeros. Thus, the location of zeros between disturbance and output critically affects {PIO} disturbance estimation, as explained in {the future work discussion in Section \ref{sec:Conclusion_FutureWork}}.

\subsection{Comparison of Discretizations with and without Induced Invariant Zeros}\label{sim5}
Consider a continuous-time second-order linear system:
\begin{equation}\label{eq:continuous}
    \begin{cases}
        \dot{x}(t) =\Phi x(t)+ G u(t)+ H f(t)\\
        y(t) = Cx(t)
    \end{cases}
\end{equation}
where \(\Phi=\begin{bmatrix}
    0 & 1\\
    0 & 0
\end{bmatrix}\), \(G=\begin{bmatrix}
    0\\
    g
\end{bmatrix}\), \(H=\begin{bmatrix}
    0\\
    1
\end{bmatrix}\), and \(C=\begin{bmatrix}
    1 & 0
\end{bmatrix}\). Discretizing the continuous-time system \eqref{eq:continuous} by zero-order hold (ZOH) yields a discrete-time system \eqref{eq_1} with the following coefficient matrices
\begin{equation}\label{eq:withzero}
    A_0=\begin{bmatrix}
    1 & T_s\\
    0 & 1
\end{bmatrix}, B_0=\begin{bmatrix}
    T_s^2g/2\\
    T_s g
\end{bmatrix}, E_0=\begin{bmatrix}
    T_s^2/2\\
    T_s
\end{bmatrix}, C_0=\begin{bmatrix}
    1 & 0
\end{bmatrix},
\end{equation}
where \(T_s\) denotes the sampling time. Note that an invariant zero at \(-1\) exists between the disturbance and the output in \eqref{eq:withzero}. By applying the Euler method, the continuous-time system \eqref{eq:continuous} can be discretized to yield \eqref{eq:withoutZero}, which does not introduce zero invariants between the disturbance and the output \cite{miklosovic2006discrete}:
\begin{equation}\label{eq:withoutZero}
    A_0=\begin{bmatrix}
    1 & T_s\\
    0 & 1
\end{bmatrix}, B_0=\begin{bmatrix}
    0\\
    T_s g
\end{bmatrix}, E_0=\begin{bmatrix}
    0\\
    T_s
\end{bmatrix}, C_0=\begin{bmatrix}
    1 & 0
\end{bmatrix}.
\end{equation}
MB-ESO in \eqref{eq_3} based on the augmented system \eqref{eq_2} is used to reconstruct the disturbance. 

\begin{figure}[htbp]
    \begin{center}
        \includegraphics[width=3.5 in]{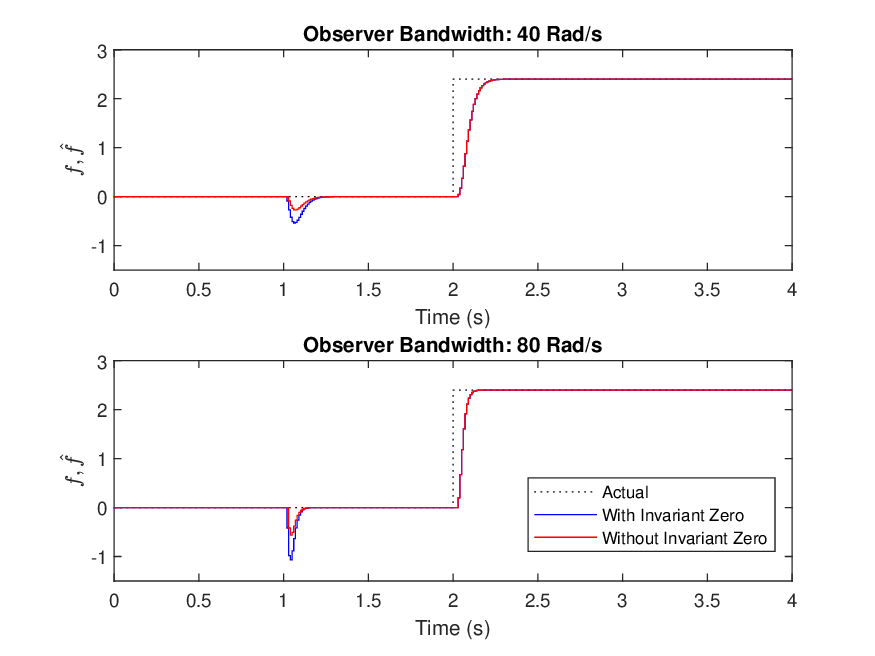}
        \caption{The reconstructed disturbances using the MB-ESO with different discretization methods and with different observer bandwidths \(\omega_o=40\) rad/s and \(\omega_o=80\) rad/s. The observer bandwidths have been converted to the continuous time domain.}
        \label{fig:discretizeWithWithoutZeros}
    \end{center}
\end{figure}

A step control input $u$ of $1$ is applied at $1$ s, followed by a step disturbance $f$ of $2.4$ at $2$ s. The control input coefficient is set to \(g=5\), and the sampling time \(T_s\) is \(0.01\)s. The initial state of the system is \(x_0=[0,0,0]^{\top}\). As illustrated in Fig. \ref{fig:discretizeWithWithoutZeros}, a disturbance estimation error occurs when the step control input is applied at \(1\) s. This error likely stems from the loss of high-frequency information during the discretization of the continuous-time system; consequently, the discrete-time MB-ESO interprets the discrepancy between the discretized and continuous-time models as a disturbance. At the instant of the control input change, the model without an invariant zero \eqref{eq:withoutZero} achieves a smaller estimation error than the model with an invariant zero \eqref{eq:withzero}. This performance degradation may be attributed to the invariant zero at \(-1\) in the MB-ESO, as predicted by the necessary and sufficient conditions proposed in this paper. Furthermore, Fig. \ref{fig:discretizeWithWithoutZeros} shows that while a higher bandwidth accelerates convergence, it also induces higher peaks in the disturbance reconstructions. Notably, both models \eqref{eq:withzero} and \eqref{eq:withoutZero} perform similarly when estimating the step disturbance at \(2\) s, where the influence of the invariant zero is less significant.


\subsection{Comparison with a Delayed Unknown Input Observer}\label{sim6}
Consider the position control of a series elastic actuator (SEA) system with two measurements~\cite{chen2021on}. The discretized SEA system with a sampling time of $20$ ms is represented by the matrices

\[
    A_0 = \begin{bmatrix}
        0 & 1 & 0 & 0\\
        -0.6587 & 1.6494 & 3.4847e-4 & 0\\
        0 & 0 & 0 & 1\\
        1.7991 & 0 & -0.9829 & 1.8929
    \end{bmatrix}, 
\]
\[B_0 = \begin{bmatrix}
        0\\
        0\\
        0\\
        74.96
    \end{bmatrix}, E_0=\begin{bmatrix}
        0 & 0\\
        0.0145 & 0\\
        0 & 0\\
        0 & 74.96
    \end{bmatrix}, C_0=\begin{bmatrix}
        1 & 0\\
        0 & 0\\
        0 & 1\\
        0 & 0
    \end{bmatrix}^{\top}.\]
It is easy to verify that this discretized SEA system has a disturbance vector relative degree \(\{2,2\}\), and 
\(V_0=\text{diag}(0.0145, 74.96)\)
is diagonal. Thus, MB-ESO in \eqref{eq_3} based on the augmented system \eqref{eq_2} is used to reconstruct disturbances. The observer gain \(L\) can be obtained using \eqref{eq:observerGain} to decouple all subsystems and place all eigenvalues at \(\omega_{o}\). 

A step input $u$ of $1$ Nm is applied at $0.1$ s, a step disturbance $f_1$ of $1.5$ Nm on the load side at $0.2$ s, and a step disturbance $f_2$ of $2$ Nm on the input side at $0.3$ s. Band-limited white noise with powers $2\times 10^{-8}$ and $2\times 10^{-1}$ is added to the first and second output measurements in both the D-UIO and MB-ESO. The plant initial state is $x_0 = [0,0,0,0]^{\top}$. For a more direct theoretical comparison, the plant models in these simulations are discretized, in contrast to the continuous-time models used in Simulation \ref{sim5}. 

\begin{figure}[htbp]
    \begin{center}
        \includegraphics[width=3.5 in]{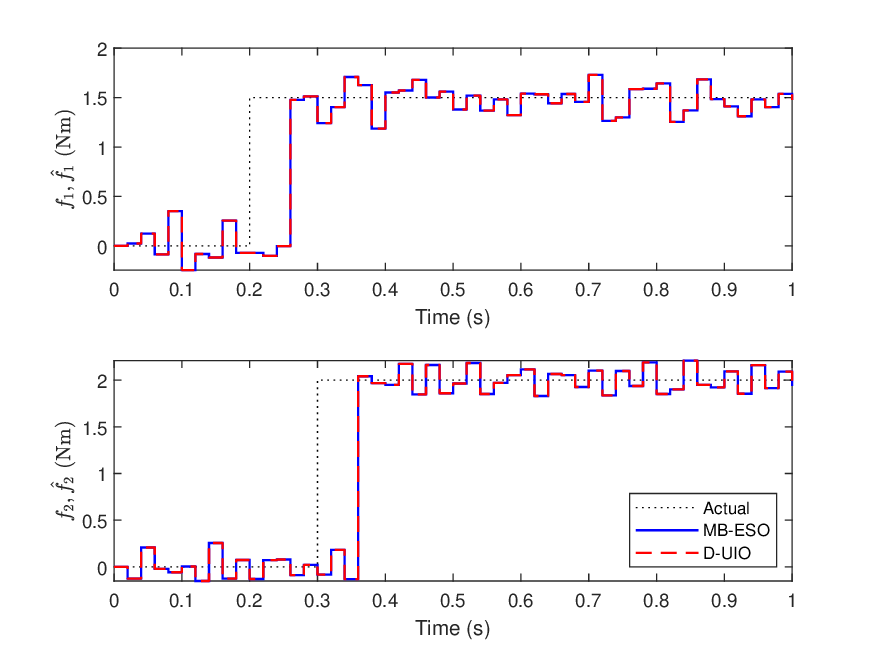}
        \caption{The true disturbances and their reconstructions by the MB-ESO and the {D-}UIO.}
        \label{fig:ESO_UIO_noise}
    \end{center}
\end{figure}

Fig.~\ref{fig:ESO_UIO_noise} shows the true and reconstructed disturbances. Due to measurement noise, the reconstructions deviate from the true disturbances. Nevertheless, the MB-ESO with eigenvalues placed at zero (infinite bandwidth) and the D-UIO with eigenvalues placed at $0.4493$ (\(-40\) rad/s in the continuous-time domain, low bandwidth) produce identical reconstructions under the same measurement noise, since both interpret noise-corrupted measurements as accurate and reconstruct disturbances that drive the measurements to those noise-corrupted values.

Moreover, the MB-ESO can smooth the reconstruction of disturbances due to the disturbance derivative term in the observer error dynamics. In motion control, the sampling time is usually \(1\) ms. After discretizing the SEA with this sampling time, the discretized system is given by 
    \[
        A_0 = \begin{bmatrix}
            \begin{smallmatrix}
                0 & 1 & 0 & 0\\
                -0.9793 & 1.9793 & 1.056e-6 & 0\\
                0 & 0 & 0 & 1\\
                0.0046 & 0 & -0.9991 & 1.9989
            \end{smallmatrix}
        \end{bmatrix}, 
        C_0 = \begin{bmatrix}
        1 & 0 & 0 & 0\\
        0 & 0 & 1 & 0
    \end{bmatrix},
    \] 

\[
    B_0 = \begin{bmatrix}
        0\\
        0\\
        0\\
        0.1904
    \end{bmatrix}, E_0 = \begin{bmatrix}
        0 & 0\\
        4.3984e-5 & 0\\
        0 & 0\\
        0 & 0.1904
    \end{bmatrix}.  
\]

\begin{figure}[htbp]
    \begin{center}
        \includegraphics[width=3.5 in]{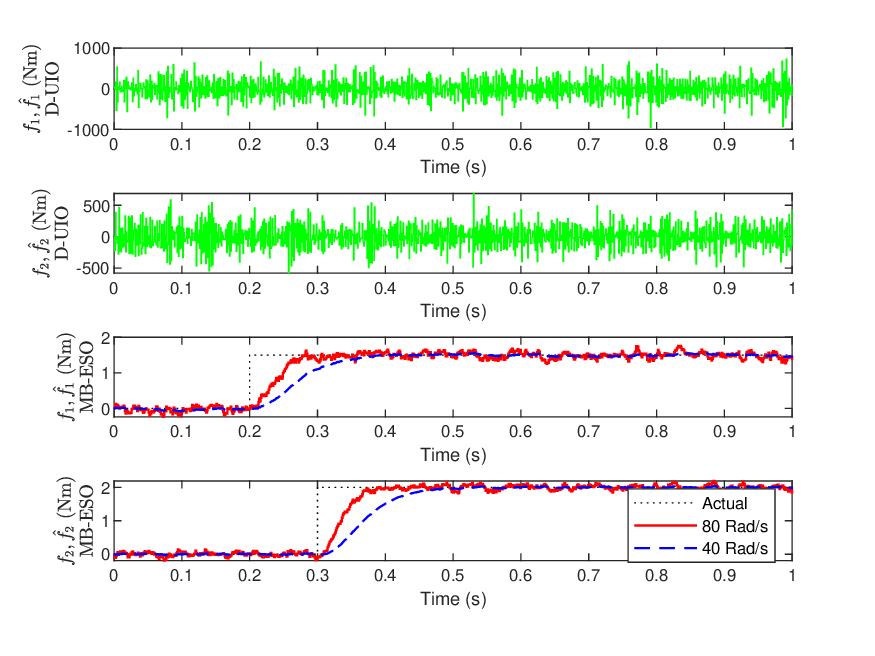}
        \caption{The reconstructed disturbances using the D-UIO and the MB-ESO with different observer bandwidth \(\omega_o=40\) rad/s and \(\omega_o=80\) rad/s. The observer bandwidths have been converted to the continuous time domain.}
        \label{fig:UIO_ESO_noise_1st}
    \end{center}
\end{figure}

The first two subfigures in Fig.~\ref{fig:UIO_ESO_noise_1st} show that the D-UIO with eigenvalues placed at \(0.9608\) (\(-40\) rad/s in the continuous-time domain, low bandwidth) produces large disturbance reconstructions because its observer error dynamics contains no uncertainty term, forcing it to track the same noisy measurements accurately under a very small sampling time. In contrast, the MB-ESO can smooth disturbance reconstructions by tuning the observer bandwidth \(\omega_o\) due to the uncertainty term in its observer error dynamics. As shown in the last two subfigures of Fig.~\ref{fig:UIO_ESO_noise_1st}, a lower bandwidth yields smoother but slower tracking of the true disturbance.  {These simulation results demonstrate that severe sensitivity to measurement noise is not driven directly by an elevated observer gain, but rather by the observer's mathematical capacity to constrain its state trajectories to replicate noisy measurement profiles, as detailed in the future work discussion in Section \ref{sec:Conclusion_FutureWork}.}


\section{Conclusion {and Future Work}}
\label{sec:conclusion}

\subsection{Conclusion}
In this paper, the novel framework of MB-ESO/GMB-ESO is proposed to exploit all available model and measurement information for improved disturbance estimation. Necessary and sufficient existence conditions are established with the only assumption that the relative degree is given between outputs and disturbances. The transient estimation properties are characterized and practical guidance for observer design is obtained. For instance, (1) the existence condition offers insight into better measurement placement for disturbance estimation; (2) in discretizing continuous-time ESOs for digital implementation \cite{miklosovic2006discrete}, additional zeros between disturbances and measurements may appear and degrade disturbance estimation performance; (3) the explicit estimation error bound is valuable for applications such as robust control barrier functions in safe control \cite{chen2023robust}; and (4) the D-UIO with no uncertainty in its observer error dynamics is compared as a benchmark with the MB-ESO to discover the mechanism behind high sensitivity to high-frequency measurement noise.



{ 
\subsection{Future Work}\label{sec:Conclusion_FutureWork}
A systematic comparison of UIOs and ASOs is provided in Table~\ref{tab:relatedWork}, and their distinct behaviors are illustrated in Simulations~\ref{sim3}, \ref{sim4}, and \ref{sim6}. Using concepts from the geometric approach, including the almost-invariant high-gain framework in \cite{willems1981almost,willems1982almost} and the weakly unobservable subspace formulation in \cite{li2026geometric}, we provide a preliminary interpretation of the mechanisms underlying UIOs and ASOs and use this perspective to explain the simulation results. Developing a rigorous and unified geometric framework that bridges these observer classes remains an important direction for future research. The preliminary insights are summarized below.

\begin{figure}[htbp]
  \centering 
  
  \begin{subfigure}[b]{0.45\textwidth}
    \centering
    \includegraphics[width=\textwidth]{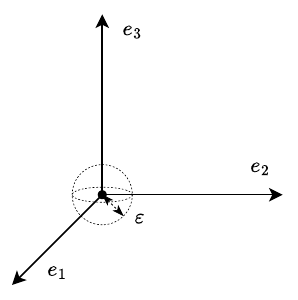} 
    \caption{No invariant zeros: the weakly unobservable subspace collapses to an empty set. } 
    \label{fig:MBESO_Geo_a}
  \end{subfigure}
  \hfill 
  \begin{subfigure}[b]{0.45\textwidth}
    \centering
    \includegraphics[width=\textwidth]{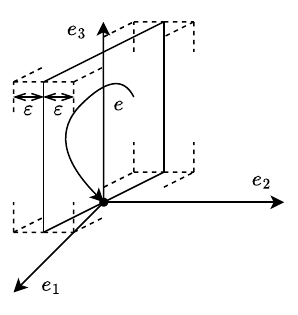} 
    \caption{Presence of invariant zeros: the weakly unobservable subspace is non-empty.} 
    \label{fig:MBESO_Geo_b}
  \end{subfigure}

  \caption{{Geometric view of the estimation errors of UIOs and ASOs, where \(\varepsilon\) is the high-gain parameter that controls the relaxation.}}
  \label{fig:MBESO_Geo}
\end{figure}

\begin{enumerate}
    \item \textit{The UIO/D-UIO Mechanism (Fig. \ref{fig:MBESO_Geo} without \(\varepsilon\)-relaxation due to its exact disturbance decoupling):} For systems without invariant zeros, the reconstruction error is confined to the origin, as shown in Fig.~\ref{fig:MBESO_Geo_a}. For systems with stable invariant zeros, the reconstruction error trajectory is confined to the weakly unobservable subspace. Because no uncertainty term perturbs the error dynamics, the error trajectory within this subspace converges asymptotically to the origin and remains at zero after convergence, as illustrated by the trajectory \(e\) in Fig.~\ref{fig:MBESO_Geo_b}. However, the UIO has no high-gain tuning parameter \(\varepsilon\) with which to adjust its tracking accuracy and therefore continuously operates at its maximum tracking capability. As a result, it is exceptionally sensitive to measurement noise, which directly explains the poor robustness of the D-UIO observed in Simulation~\ref{sim6}.

    \item \textit{The PIO/MB-ESO High-Gain Mechanism (Fig. \ref{fig:MBESO_Geo} with \(\varepsilon\)-relaxation):} Because the PIO and MB-ESO belong to the high-gain observer paradigm, their behavior is governed by an adjustable relaxation parameter \(\varepsilon\). For systems without invariant zeros, although an uncertainty term perturbs the observer error dynamics, the high-gain mechanism bounds the error trajectory and forces the estimation error to converge to a ball of radius \(\varepsilon\), as shown in Fig.~\ref{fig:MBESO_Geo_a}. As the observer gain increases, this ball can be made arbitrarily small. Moreover, the chain-of-integrator form widely used in ESOs, EHGOs, and GPIOs is a special structure with no invariant zeros.
    
    In contrast, for systems with invariant zeros, the observer provides no active structural mechanism for driving the error toward the origin within the weakly unobservable subspace, as illustrated in Fig.~\ref{fig:MBESO_Geo_b}. This also explains why the boundedness assumptions commonly imposed in the literature are insufficient: they cannot eliminate the reconstruction ambiguity associated with the weakly unobservable subspace. Consequently, the only mechanism that drives the error trajectory toward the origin is the internal stability of the invariant zeros associated with this subspace.
    
    This behavior differs fundamentally from that of the UIO framework, whose error dynamics are completely free of disturbance-derivative perturbations and whose reconstruction error remains at the origin after convergence. This mechanism explains the performance variations observed in Simulation~\ref{sim4}, where the locations of the invariant zeros strongly affect reconstruction performance. For example, if the invariant zeros are strongly stable, that is, close to the origin in the discrete-time domain, and the observer bandwidth is kept relatively low, so that \(\varepsilon\) remains comparatively large, the error envelope in Fig.~\ref{fig:MBESO_Geo_b} can closely resemble the zero-free case in Fig.~\ref{fig:MBESO_Geo_a}. This explains why the PIO can still achieve disturbance reconstruction with limited oscillation in certain cases, as illustrated in Simulation~\ref{sim4}, while failing to provide a general convergence guarantee without the structural condition proposed in this work.

\end{enumerate}

}












\bibliographystyle{unsrtnat}

\bibliography{cas-refs}



\end{document}